\documentclass{ws-ijbc}
\usepackage{ws-rotating}     % used only when sideways tables/figures are used
\usepackage{graphicx}
\usepackage{epstopdf}

\usepackage{needspace} % opcional, mas ajuda

\usepackage{colortbl}
\usepackage[dvipsnames]{xcolor}

\usepackage{hyperref}

\begin{document}

\catchline{}{}{}{}{} % Publisher's Area please ignore

\markboth{Franco, F.F. and Paula, G.T. and Chertovskih, R. and Oliveira, D.N. and Rempel, E.L.}{Transition to chaos in two-dimensional Rayleigh-B\'enard convection: the role of the magnetic field}

%\title{INSTRUCTIONS FOR TYPESETTING MANUSCRIPT\\
%USING \LaTeX\footnote{For the title, try not to use more than
%three lines. Typeset the title in 15 pt Times Roman, uppercase and
%boldface.}}

\title{Transition to chaos in two-dimensional Rayleigh-B\'enard convection: the role of the magnetic field}

\author{Francis F. Franco}
\address{Government of the State of Goiás (SEDUC-GO), CEPI João Roberto Moreira, 75802-124, Jataí, GO, Brazil\\
francis.franco@educa.go.gov.br}%francis\_franco@ufj.edu.br}

\author{Gabriel de T. Paula}
\address{Aeronautical Engineering Division, Technological Institute of Aeronautics (ITA), 12228-900, S\~ao Jos\'e dos Campos, SP, Brazil\\
gabrielgtp@ita.br}

\author{Roman Chertovskih}
\address{Research Center for Systems and Technologies (SYSTEC--ARISE), Faculty of Engineering, University of Porto,  4200-465, Porto, Portugal\\
roman@fe.up.pt}

\author{Dalton N. Oliveira}
\address{Federal Rural University of the Amazon (UFRA), 68515-000, Parauapebas, PA, Brazil\\
dalton.oliveira@ufra.edu.br}

\author{Erico L. Rempel}
\address{Department of Mathematics, Technological Institute of Aeronautics (ITA), 12228-900, S\~ao Jos\'e dos Campos, SP, Brazil\\
rempel@ita.br}

\maketitle

\begin{history}
\received{(to be inserted by publisher)}
\end{history}

\begin{abstract}
The impact of an externally imposed magnetic field on numerical simulations of two-dimensional Rayleigh-B\'enard convection (RBC) is investigated. Initially, the RBC model is examined in the absence of a magnetic field to establish a baseline. Then, a background magnetic field is introduced, and its influence on the transition to chaos is explored. For the purely hydrodynamic case and a range of the reduced Rayleigh number, the system exhibits traveling rolls which,  
after an attractor-merging crisis, give way to chaotic traveling rolls. Upon imposing a background magnetic field, there is a notable increase in the occurrence of traveling roll dynamics. Furthermore, the presence of the magnetic field favors the splitting/breaking of convective rolls, indicating a possible mechanism for transition to two-dimensional turbulence, with the structure of the convection cell being disrupted. %\textcolor{red}{The origin of this breaking of convection rolls is due to a chaotic saddle in the background, which presents this behavior.}
 A detailed analysis of the velocity field reveals that the collision between a saddle point and the center of a convective roll restores the system's original topology, with two symmetric kinetic vortices. During this collision, a magnetic vortex splits in two as a result of a magnetic reconnection. This behavior occurs intermittently in time.

\end{abstract}


%\begin{multicols}{2}
\section{Introduction}
Magnetoconvection concerns the interaction between convective motions in conducting fluids and applied magnetic fields \cite{chandra}. In this scenario, the fluid is strongly governed by the interaction between gravitational buoyancy and the Lorentz force \cite{mccormack2025}. This phenomenon is present in several highly relevant contexts; for example, in astrophysics, stars such as the Sun have convective zones, in which rotation, convection, and magnetic fields interact continuously \cite{yutobekki2025}. A similar phenomenon occurs in the interior of the Earth, where the magnetic field is generated by electric currents in the liquid core by convective flows, in the so-called planetary dynamo \cite{Schaeffer2017}. Magnetoconvection is also relevant in industrial applications, e.g., \citet{akhtari2024} highlights the importance of magnetoconvection in the fabrication of large semiconductor crystals, in which the Czochralski method is applied for crystal growth \cite{muller1988}; in the fabrication of high-quality steels \cite{cukierski2008}; and in the operation of liquid metal batteries \cite{mistrangelo2021}. Furthermore, \citet{mccormack2025} highlights the importance of this physical phenomenon in liquid-metal cooling systems of nuclear fusion reactors. \citet{macek2010} further exemplify applications of magnetoconvection, citing, for example,  applications in magnetically confined plasmas in tokamaks. In this context, \citet{wilczynski2019} show that the plasma dynamics in the scrape-off layer of tokamaks can be interpreted as a natural convection problem with additional effects.

Given the importance of these systems, mathematical models play a fundamental role in understanding and predicting their behavior.  In this context, the Rayleigh-B\'enard convective (RBC) model stands out \cite{rayleigh1916}, an idealized version of thermal convection in which a fluid is contained between two horizontal plates separated by a height ($h$), where the lower plate has a higher temperature than the upper plate, generating a temperature gradient $(\delta T)$. The two-dimensional RBC remains of great interest because it is simpler to study, precisely due to the geometry and thermal properties involved, while still retaining the essential physics present in the geophysical, astrophysical, and engineering systems discussed previously \cite{hepworth2014}. The RBC model with periodic boundary conditions in \(x\) and \textit{free-slip} conditions in \(z\) has been widely used to study instabilities, as in \citet{winchester2022}, who analyze the stability of stationary rolls by varying the Prandtl number (\(Pr\)) and the Rayleigh number (\(Ra\)); convective modes, as in \citet{Paul2012}, who investigate several dynamical transitions using bifurcation diagrams generated by direct numerical simulation (DNS); and turbulence, as in \citet{goluskin2014}, where the authors investigate, through DNS, how strong horizontal turbulence can reduce vertical heat transport. \citet{wang2020} further highlight that RBC under periodic and free-slip boundary conditions has been widely used as a primary model for the study of zonal flows.

When the fluid under consideration is electrically conducting and an external magnetic field is applied, the problem becomes one of magnetoconvection. The two-dimensional magnetoconvection model is usually defined with periodic boundary conditions in $x$, \textit{free-slip} conditions in $z$, and, with the addition of the magnetic field, the condition $\partial A/\partial z = 0$ at the plates. These boundary conditions are adopted because, according to \citet{Weiss2014} (p.~55), they are mathematically self-consistent and convenient.  \citet{rucklidge1996} investigated the nonlinear dynamics of this system, studying the behavior of the equations, reduced-order models, and maps in both magnetic and non-magnetic cases, and investigating how the symmetry of periodic solutions changes as a result of global bifurcations.   \citet{Dawes2007} investigated localized structures, showing that, in the presence of a vertical magnetic field, convection can organize itself into localized cells. This line of research was extended by the studies of \citet{lojacono2011,lojacono2012}, who investigated these localized structures through numerical simulations with fixed and varying Chandrasekhar number $Q$. Taken together, these studies show that this two-dimensional formulation can capture several essential nonlinear mechanisms of magnetically constrained convection. While three-dimensional models have successfully employed dynamical systems theory and Lyapunov spectra to map complex routes to weak turbulence and hyperchaos \cite{Roman2015}, the two-dimensional formulation allows for a more tractable isolation of topological dynamics and magnetic reconnections.

The objective of this work is to better understand the two-dimensional magnetoconvection system in the transition-to-chaos regime and the role of the magnetic field. The analysis is conducted by varying two control parameters: the Chandrasekhar number $Q$, associated with the background magnetic field $\boldsymbol{B}_0$, and the Rayleigh number $R_a$, associated with the temperature gradient $\delta T$.
Two main scenarios were explored: \(Q = 0\) and \(Q > 0\). In the purely hydrodynamic convection case (\(Q = 0\)), a bifurcation diagram is generated as a function of the reduced Rayleigh number \(r\), defined as $r = R_a / R_c$, where \(R_c \approx 657.51\) is the critical value for the onset of convective motions \cite{chandra}. %A Ruelle-Takens  route to chaos \cite{ruelle1971} is observed. In addition, a hysteresis scenario associated with multistability was demonstrated. For the magnetohydrodynamic case \(Q > 0\), it was found that the presence of a magnetic field facilitates the appearance of chaotic traveling rolls and spatiotemporal chaos. The magnetic field also leads to the
% breakup of the convective rolls. 
To analyze the evolution of spatial patterns in this process, an approach based on the Finite-Time Lyapunov Exponent (FTLE) \cite{haller2000,haller2001,haller2015} is applied to both the velocity and the magnetic fields. In addition, the Instantaneous Vorticity Deviation (IVD) \cite{haller2016} is used to find vortices in the velocity field, and the Local Current Deviation (LCD) \cite{rempel2016,rempel2019} for the magnetic field. %This analysis made it possible to identify that, during the breakup process, a saddle point emerges in the velocity field, associated with the formation of an eight-shaped homoclinic orbit \cite{gonchenko13}. In addition, it was observed that, after the breakup, when the system recovers its initial symmetry, a duplication of magnetic vortices occurs in the magnetic field in the region close to the one where, in the velocity field, the collision of the saddle point with one of the new centers formed during the breakup occurs.

The article is organized as follows: Section 2 presents the nonlinear tools used in the analysis: the FTLE, IVD, and LCD. Section 3 presents the two-dimensional model and its numerical solution. Section 4 presents the results obtained. Finally, Section 5 provides the main conclusions of the work.

\section{Nonlinear tools}

%----------------%-----------------%------------------%----------------%
\subsection{Finite Time Lyapunov Exponent (FTLE)}
\label{sec ftle}

Usually, the FTLE fields are computed by accounting for the temporal variations of the velocity field to study the motion of fluid particles and detect transport barriers \cite{haller2000,haller2015,rempel2013}. Here, we compute each FTLE field for a stationary velocity field snapshot because we are not interested in fluid transport, but in detecting instantaneous invariant structures in the velocity field. Thus, the FTLE fields are computed as
\begin{equation}
\sigma^{\xi}(\boldsymbol{x}):=\frac{1}{|\xi|}\log\sqrt{\lambda},
\label{eq ftle}
\end{equation}
where $\xi$ is a finite scale and $\lambda$ is the largest eigenvalue of the right Cauchy-Green deformation tensor $C^{\xi}(\boldsymbol{x}) = (\nabla F^{\xi}(\boldsymbol{x}))^T\nabla F^{\xi}(\boldsymbol{x})$, where $T$ denotes the transposed operation and $\nabla F^{\xi}(\boldsymbol{x})$ is the Jacobian matrix of the flow, i.e. the deformation gradient tensor computed for $\xi$ ``time'' units. The position vector $\boldsymbol{x}(s):[0,\xi]\rightarrow \mathbb{R}^2$ is obtained as a solution to the initial value problem

\begin{equation}
\frac{\mbox{d}\boldsymbol{x}}{\mbox{d}s} = \boldsymbol{v}(\boldsymbol{x}(s),t_0),\qquad \boldsymbol{x}(0) = \boldsymbol{x}_0,
\end{equation}
where the integral curve has been parameterized by a parameter $s$ and we have fixed the vector field $\boldsymbol{v}$ at instant $t_0$.

%----------------%-----------------%------------------%----------------%
%\subsection{Vortex detection methods}
%\label{sec vortex}

\subsection{Instantaneous Vorticity Deviation (IVD)}

Consider a velocity field $\boldsymbol{v}(\boldsymbol{x},t)$ in a two-dimensional fluid domain $U(t)$. The vorticity is defined as the curl of the velocity, $\boldsymbol{\omega}(\boldsymbol{x},t) = \nabla \times \boldsymbol{v}(\boldsymbol{x},t)$. At a given instant $t$, we can compute the spatially averaged vorticity over the domain considered, denoted by 
\begin{equation} \label{EQ:vorticidade_media_espacial}
\overline{\boldsymbol{\omega}}(t) = 
\frac{
    \displaystyle\int_{U(t)} \boldsymbol{\omega}(\boldsymbol{x}, t) \, \mathrm{d}A
}{
    \mathcal{A}\left( U(t) \right)
} \, ,
\end{equation}
where $\mathcal{A}(U(t))$ refers to the area of the two-dimensional domain and \( \mathrm{d}A \) denotes an area element. The \textit{Instantaneous Vorticity Deviation} (IVD) is defined as the difference between the local vorticity and the mean vorticity  \cite{haller2016}

\begin{equation}
\text{IVD}(\boldsymbol{x}, t) = \big\|\,\boldsymbol{\omega}(\boldsymbol{x}, t) - \overline{\boldsymbol{\omega}}(t)\,\big\|, \
\label{eq:IVD}
\end{equation}
where $\|\cdot\|$ denotes the Euclidean norm of the vector. %In words, $\text{IVD}(\mathbf{x},t)$ quantifies how much the vorticity at that point differs from the global mean rotation of the fluid at that instant. 
If the mean vorticity $\overline{\boldsymbol{\omega}}(t)$ is nonzero (for instance, if the fluid exhibits an overall/global rotation), this contribution is subtracted, ensuring that the IVD captures only relative rotations. In a flow with no global rotation ($\overline{\boldsymbol{\omega}} = \mathbf{0}$), the IVD simply coincides with the magnitude of the local vorticity $\|\boldsymbol{\omega}\|$. In any case, as shown by \citet{haller2016}, this definition has the remarkable property of being objective, i.e., invariant under arbitrary time-dependent translations or rotations. %In practical terms, this means that if we analyze the same flow field in a rotating frame, the IVD value at each point will be the same as in a fixed frame, because both $\boldsymbol{\omega}$ and $\overline{\boldsymbol{\omega}}$ are affected in the same way by the frame rotation, and their difference remains invariant.

The IVD itself is an instantaneous, Eulerian measure: at each time, it provides a scalar field indicating regions of high relative rotation. An instantaneous 2D vortex core is defined as the point where the IVD attains a local maximum. %, and the vortex boundary is defined as a convex IVD contour line around that maximum. The convexity requirement ensures that the identified region is approximately circular/elliptical, excluding, for example, elongated bands of vorticity (which typically represent shear layers rather than vortices). 
 This technique is robust in detecting rotating tubular structures and has been validated in several benchmark flows \cite{haller2016}.

\subsection{Local Current Deviation (LCD)}

By direct analogy with the IVD, defined above, an objective Eulerian magnetic vortex can be defined by replacing the vorticity (the rotation of the velocity field) with the current density field (the rotation of the magnetic field).

We therefore define the mean current over the domain at each instant as

\begin{equation}
\overline{\boldsymbol{J}}(t) = 
\frac{
    \displaystyle\int_{U(t)} \boldsymbol{J}(\boldsymbol{x}, t) \, \mathrm{d}A
}{
    \mathcal{A}\left( U(t) \right)
} \, .
\end{equation} 

The \textit{Local Current Deviation} (LCD) is given by \cite{rempel2016,rempel2019} 

\begin{equation}
\text{LCD}(\boldsymbol{x}, t) = \big\|\,\boldsymbol{J}(\boldsymbol{x}, t) - \overline{\boldsymbol{J}}(t)\,\big\| ,\
\label{eq:LCD}
\end{equation}
where $\boldsymbol{J} = \nabla \times \boldsymbol{B}/\mu_0$ is the electric current density. % (in SI units, or multiplied by convenient constants). 
 Like the IVD, the LCD measures the intensity of the local current relative to the global mean current, providing a scalar measure of the local ``magnetic twisting intensity''. Regions with high LCD values indicate where the magnetic field is significantly more rotational than the average, thereby delineating the interior of magnetic vortices.

%----------------%-----------------%------------------%----------------%
\section{The model and numerical solution}
\label{sec model}

We consider an electrically conducting incompressible two-dimensional fluid flow in a horizontal plane layer, uniformly heated from below and cooled from above, and subjected to an imposed vertical magnetic field. The system is buoyancy-driven, and the Boussinesq approximations (see, e.g., \cite{chandra}) are assumed. The governing (non-dimensional) equations of the considered convective hydromagnetic system are 
\cite{knobloch1986, Arter1983, Dawes2007}:

\begin{equation} \label{conserv_b}
\dfrac{\partial \omega}{\partial t} + J [ \psi,\omega ] = -P_{r} R_{a} \dfrac{\partial \theta}{\partial x} -P_{r} P_{m} Q  \left( J [A, \nabla^{2} A] + \dfrac{\partial \nabla^{2} A}{\partial z} \right) + P_{r} \nabla^{2} \omega,
\end{equation}
\begin{equation} \label{induzb_b}
\dfrac{\partial A}{\partial t} + J [ \psi,A ] = \dfrac{\partial \psi}{\partial z} + P_{m} \nabla^{2} A,
\end{equation}
\begin{equation} \label{difust_b}
\dfrac{\partial \theta}{\partial t} + J [ \psi,\theta ] =  \nabla^{2} \theta + \dfrac{\partial \psi}{\partial x},
\end{equation}
where $J[f,g] \equiv [(\partial f/ \partial x)(\partial g/ \partial z)-(\partial f/ \partial z)(\partial g/ \partial x)]$ \cite{Bekki1995}. The velocity field is given by 

\begin{equation}
 \boldsymbol{v} = \nabla \times [\psi(x,z,t) \boldsymbol{\hat{y}}] = \left[-\frac{\partial \psi}{\partial z}\ ,\ 0\ ,\ \frac{\partial \psi}{\partial x} \right],
\end{equation}
where $\psi$ is the stream function and the vorticity is $\omega=\nabla\times \boldsymbol{v}=-\nabla^{2} \psi \boldsymbol{\hat{y}}$. $\theta(x,z,t)$ is the difference between the temperature and the conduction profile $T=1-z$. Considering the magnetic potential $A(x,z,t)$, the magnetic field is represented as  
\begin{equation}
\boldsymbol{B}=\boldsymbol{B}_{0}+\nabla \times [A(x,z,t) \boldsymbol{\hat{y}}]= \left[-\frac{\partial A}{\partial z}\ ,\ 0\ ,\ \frac{\partial A}{\partial x} + 1 \right],
\end{equation}
where $\boldsymbol{B}_{0}=(0, 0, 1)$ is an imposed uniform vertical field. We  assume periodicity in the horizontal direction $x$ with period $L$, and the free-slip %stress-free
constant temperature upper and lower boundaries (i.e., $\psi=\omega=0$, $\partial_z A=0$ and $\theta=0$ at $z=0$ and $z=h$). The units of length and time are $h$ and $h^2/\kappa$, respectively, $\boldsymbol{v}$ is in units of $\kappa/h$, $\theta$ in units of $\Delta T$ and $\boldsymbol{B}$ in units of $\sqrt{\mu_0\rho}\kappa/h$.

\subsection{Control parameters}
 
\noindent Four non-dimensional control parameters are usually employed to describe the properties of convective flows in the presence of an imposed magnetic field~\cite{chandra, Mondal2018}: the Rayleigh number $R_{a}$, the Chandrasekhar number $Q$, the Prandtl number $P_{r}$, and the magnetic Prandtl number $P_{m}$, defined as
\begin{equation}
R_{a}= \frac{g \alpha \Delta T h^{3}}{\kappa \nu}, \qquad Q=\dfrac{|\boldsymbol{B}_{0}|^2 h^{2}}{\mu_{0} \rho_{0} \nu \eta}, \qquad P_{r}= \nu/ \kappa, \qquad P_{m}=\nu / \eta,
\end{equation}
where $g$ is the acceleration due to gravity, $\alpha$ is the thermal expansion coefficient, $\Delta T$ is the temperature difference across distance $h$, $\kappa$ is the thermal diffusivity, $\nu$ is the kinematic viscosity, $\boldsymbol{B}_0$ is the (imposed) background magnetic field, $\mu_0$ is the vacuum magnetic permeability, $\eta$ is the magnetic diffusivity and $\rho_0$ is the mean mass density. These parameters can vary significantly for geophysical and astrophysical flows. For example, in the Earth's liquid outer core, the Prandtl number is estimated to be of the order of 0.1, whereas in the solar convective zone, the Prandtl number is extremely low ($P_{r} \sim 10^{-8}$)~\cite{Thual1992}. As a matter of comparison, $P_{r}=0.02$ for mercury, $P_{r}=0.01$ for liquid sodium \cite{Paul2012}. The magnetic Prandtl number in the Earth's outer core is $P_{m} \sim 10^{-6}$ ~\cite{Schaeffer2017, Mondal2018} and for stellar interiors it can vary from $10^{-6}$ to $10^{-4}$ \cite{Mondal2018}. 
Regarding the Chandrasekhar number, it can reach $~10^{15}$ in the Earth's outer core~\cite{Yan2019}, but much smaller values are usually adopted in laboratory experiments and numerical simulations, where $Q\leq 10^{6}$ \cite{Cioni2000, Yan2019, Reddy2020}. 

In Rayleigh-B\'enard convection, the value of the Rayleigh number for the Earth's outer core is approximately $10^{23}$ ~\cite{Christensen2006, Jones2007}. If the Rayleigh number exceeds a critical value, $R_c$, a convective motion sets in (i.e., the trivial motionless state is unstable to perturbations). In the convective system with free slip (also called stress-free) horizontal boundaries of the layer, the critical value is~\cite{chandra}
\begin{equation}
R_{c}=27 \pi^{4}/4 \approx657.51.
\end{equation}
In what follows, we use the reduced Rayleigh number, defined as $r=R_{a}/R_{c}$.

Unfortunately, it is computationally too heavy to perform numerical simulations for realistic values (e.g., for the Earth's outer core or the solar convective zone) of $P_{r}$, $P_{m}$, $R_{a}$, and $Q$, because of the huge numerical resolution required. In the present work, the parameters are chosen based on the literature on laboratory and numerical experiments of RBC and magnetoconvection. The chosen Prandtl number is $P_{r}=6.8$, following \cite{Paul2012} and \cite{Podv2008}. The chosen magnetic Prandtl number is $P_{m}=1$ \cite{knobloch1986, Proctor1982, Bekki2007, Bovino2013, Weiss2014}. The reduced Rayleigh number $r$ was explored inside the interval $1 \leq r \leq 1000$ and the Chandrasekhar number inside $0 \leq Q \leq 100$, the same interval for $Q$ was explored by \cite{Arter1983} and \cite{Weiss2014} in the study of magnetoconvection, and \cite{Bushby2005} in the study of intermittency in the solar photosphere.

Although realistic geophysical and astrophysical plasmas typically operate at $P_m \ll 1$, adopting $P_m = 1$ serves as a fundamental benchmark, where viscous and magnetic diffusion timescales are strictly balanced. This choice allows us to clearly isolate the purely topological dynamics of magnetic reconnections and vortex interactions, decoupled from the extreme boundary-layer separation and numerical stiffness inherent to low-$P_m$ regimes.

\subsection{Numerical solutions}

Equations (\ref{conserv_b})-(\ref{difust_b}) are solved numerically using the standard pseudospectral method \cite{canuto06} in a rectangle of size [L,1] with $L=2\sqrt{2}$. The fields $\omega(x,z,t), A(x,z,t)$ and $\theta(x,z,t)$ are represented as Fourier series in all spatial variables (complex exponentials in the horizontal direction and sine/cosine in the vertical direction), derivatives are computed in the Fourier space, multiplications are performed in the physical space, and the Orszag 2/3-rule is applied for dealiasing. The system of ordinary differential equations for the Fourier coefficients is solved using the third-order exponential time-differencing method ETDRK3 \cite{cox02}, where the linear terms originated by the Laplacians in (\ref{conserv_b})-(\ref{difust_b}) are integrated exactly. We used the fixed integration time-step $dt = 10^{-5}$.

\begin{figure}[!htp]
\begin{center}
\includegraphics[width=\columnwidth]{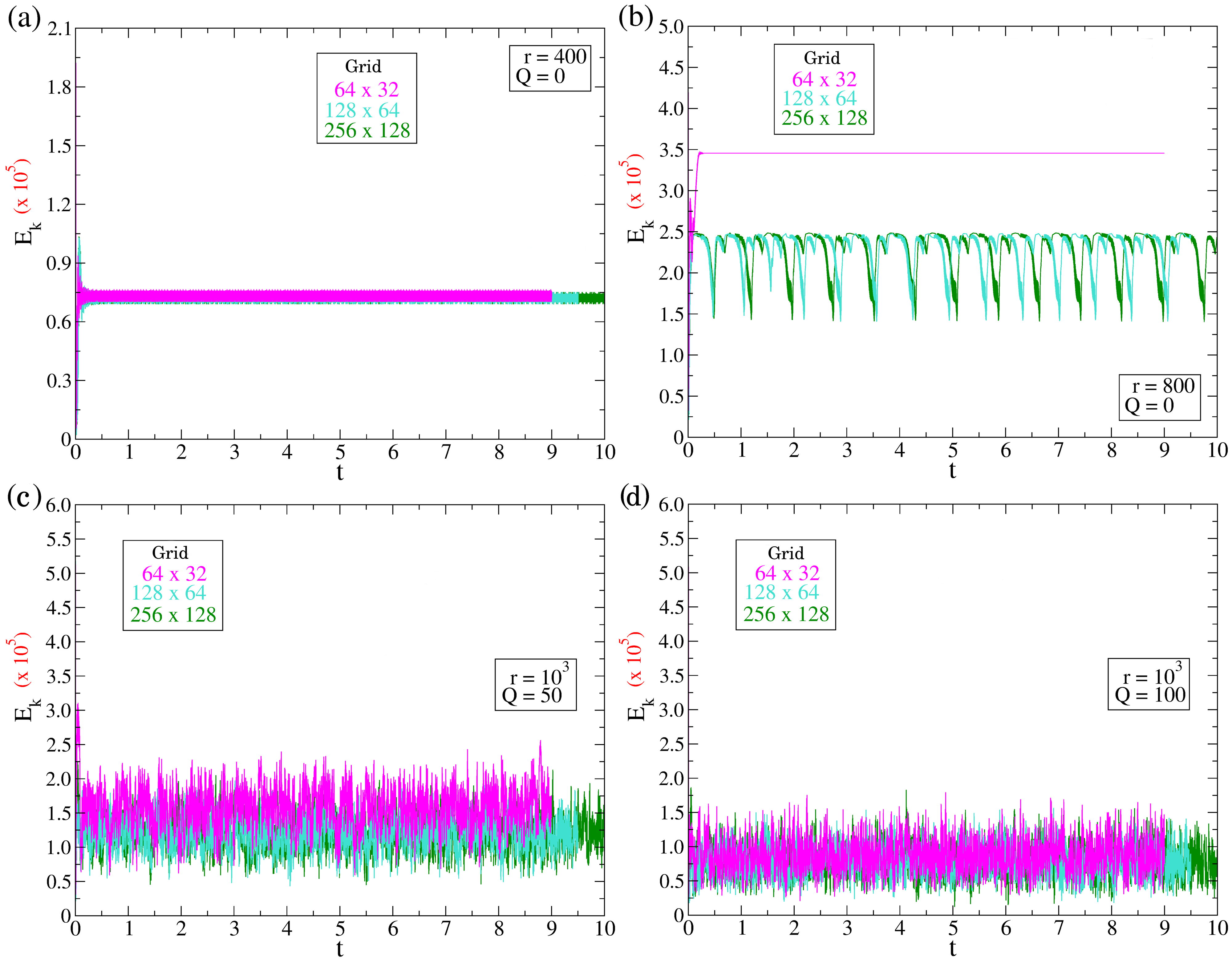}
\end{center}
\caption{Test of numerical resolution. The time series of the kinetic energy is shown
for three different resolutions and different choices of $r$ and $Q$, as indicated in each panel.}
\label{Fig1}
\end{figure}

We set $P_r = 6.8$ and $P_m = 1$ and vary $Q$ and $r$ as control parameters. To determine appropriate numerical resolution, the kinetic energy is computed at different grid resolutions, corresponding to different numbers of Fourier modes in the pseudospectral method. Figure \ref{Fig1} shows the results for three different resolutions, low resolution ($64 \times 32$, magenta), mid resolution ($128 \times 64$,,  cyan) and high resolution ($256 \times 128$, dark green). Figure \ref{Fig1}(a), displays the time series of the kinetic energy for $r=400$ and $Q=0$; all three systems exhibit periodic behavior with a similar energy level. Figure \ref{Fig1}(b) displays the time series of the kinetic energy for $r=800$ and $Q=0$; in the low-resolution system, an attractor with steady kinetic energy emerges, while in both mid and high-resolution systems, attractors with lower energy and of oscillatory behavior in time are found. In Fig. \ref{Fig1}(c), for $r=1000$ and $Q=50$, all three systems exhibit chaotic oscillations with a similar energy level, but the low resolution still has a slightly higher amplitude. When the imposed magnetic field is increased, the kinetic energy tends to decrease, and attractors for the three resolutions show the same behavior, as seen in Figs. \ref{Fig1}(c) and (d), for $r=1000$ and $Q=50$ and $r=1000$ and $Q=100$, respectively. We performed several other similar tests and based on the results, we adopted the mid-resolution for the hydrodynamic (in the absence of magnetic field) simulations discussed below. For the magnetohydrodynamic simulations (in the presence of a magnetic field) sometimes a higher resolution is employed. To validate our numerical tools, in the next section we present a comparison with the results obtained by \citet{Paul2012}.

\section{Nonlinear Dynamics Analysis}
\label{sec analysis}

\subsection{2D Rayleigh-B\'enard Convection} \label{dbr}
\begin{figure}[!htp]
\begin{center}
\includegraphics[width=0.5\columnwidth]{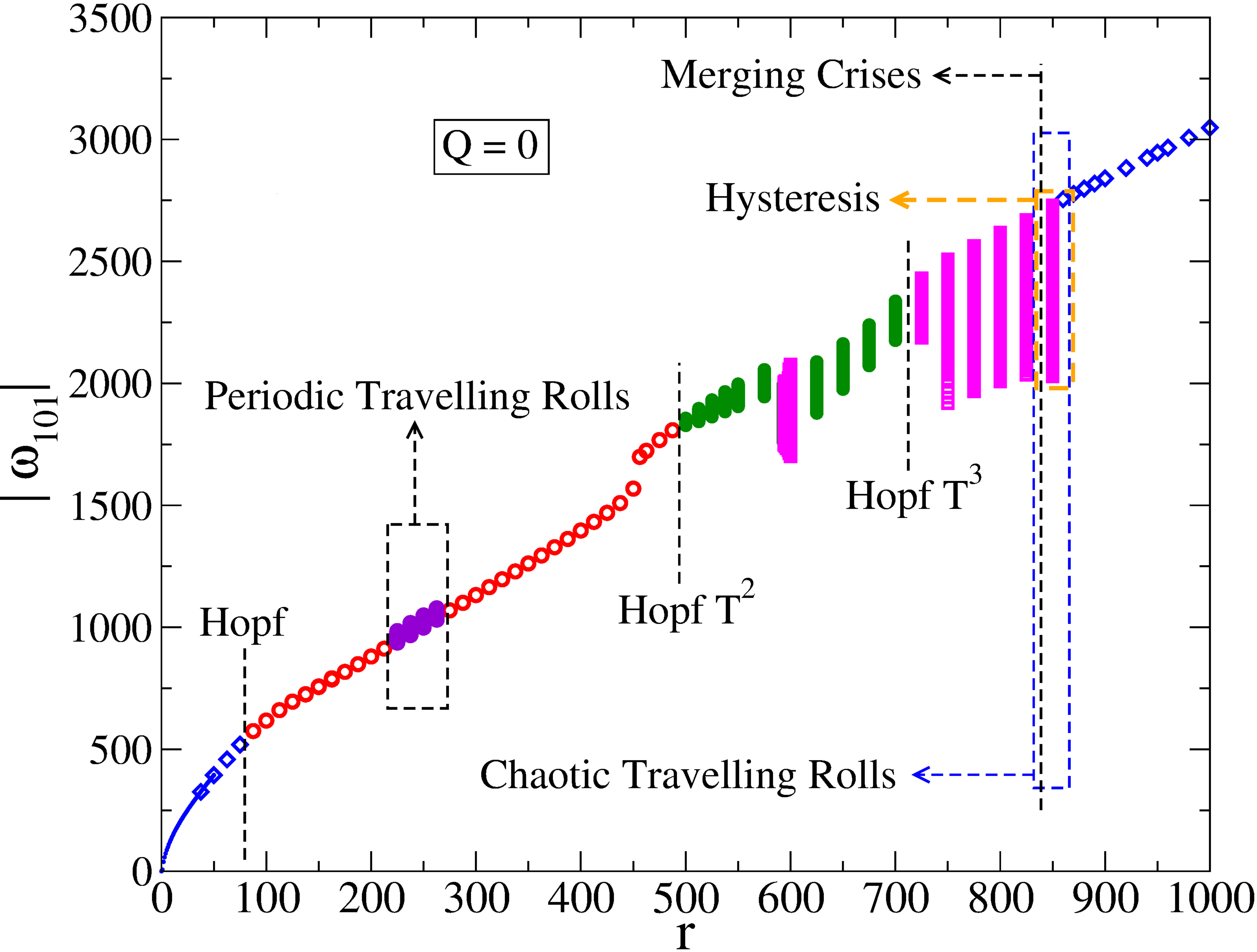}\end{center}
\caption{Bifurcation diagram for $|\omega_{101}|$ as a function of $r$. Blue diamonds represent fixed point attractors, red circles denote periodic attractors, purple circles are periodic traveling rolls, green circles are quasiperiodic attractors, and magenta squares represent chaotic attractors.}
\label{Fig2}
\end{figure}

In this section, we reproduced the results of \citet{Paul2012}, where the purely hydrodynamic system (in the absence of magnetic field) was studied. We computed a bifurcation diagram as a function of the reduced Rayleigh number $r$, as shown in Fig. \ref{Fig2}. For each value of $r$, we plot the magnitude of the most energetic Fourier mode of the vorticity ($|\omega_{101}|$) from the corresponding attractor (i.e., the initial transients are dropped). With the exception of the fixed point behavior at the left branch of the diagram, all points represent the sequence of maxima of oscillatory time series of the mode. In the diagram, blue diamonds represent fixed point ($A^{FP}$) behavior, red circles represent periodic attractors ($A^{P}$); purple circles denote periodic traveling rolls attractors ($A^{PTR}$); green circles refer to quasiperiodic attractors ($A^{QP}$) and magenta squares denote chaotic attractors ($A^{C}$). A fixed point (steady) attractor exists for $0<r<87$; then the system undergoes a Hopf bifurcation at $r\approx 87$ and a periodic attractor $A^{P}$ is created (an attracting, time-periodic limit cycle). The $A^{P}$ is stable for $87<r<219$; as $r$ increases beyond 220, the emerging attractor $A^{PTR}$ is a periodic traveling rolls (PTR) behavior. The nature of such attractors is discussed in the next paragraph. The PTR is stable up to $r\approx 275$, when the emerging attractor is again a time periodic one, existing up to $r\approx 500$. Then, a secondary Hopf bifurcation takes place, giving rise to a quasiperiodic attractor $A^{QP}$ (a torus $T^2$ originated by a trajectory with two incommensurate temporal frequencies). Following the Ruelle-Takens route to chaos, at $r\approx 593$ the system undergoes a transition from quasiperiodicity to chaos. The chaotic attractor is stable until $r\approx 600$, and for $600<r<725$ the system returns to a quasiperiodic attractor. At $r\approx 725,$ a structurally unstable torus $T^3$ is formed, and for $725<r<858,$ the attractors are chaotic $A^{C}$. In the final part of this interval, we observed chaotic traveling rolls and a merging crisis, which are discussed below. For $858<r<1000,$ the attractor is again a fixed point. Our simulations reproduced all regimes and bifurcations described in the bifurcation diagram shown in Fig. 3 of \citet{Paul2012}, although they plotted the magnitude of the Fourier mode of $v_z$ rather than $\omega$. There is a tiny discrepancy in the values of $r$ at which local and global bifurcations occur; this may be due to different numerical parameter values used in the simulations. Below, we describe in more detail some of the regimes and bifurcations that are of interest for the analysis of the system in the presence of magnetic field.

\begin{figure} [htp!]
 \centering
 \includegraphics[width=\textwidth]{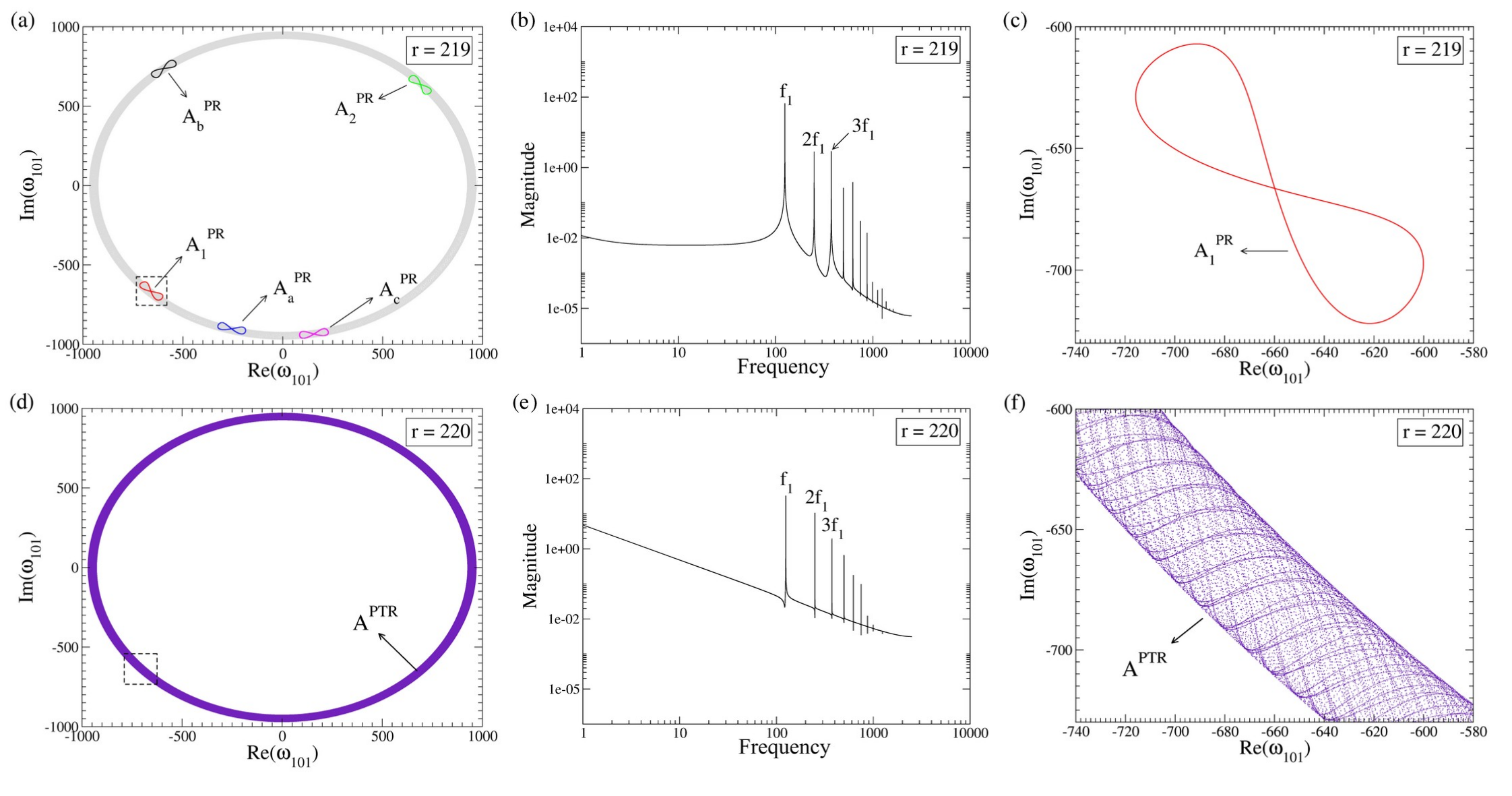}
 \caption{Phase space projection on $Im(\omega_{101}) \times Re(\omega_{101})$. (a) five symmetric periodic attractors: ${A_{a}}^{PR}$ (blue), ${A_{b}}^{PR}$ (black), ${A_{c}}^{PR}$ (magenta), ${A_{1}}^{PR}$ (red) and ${A_{2}}^{PR}$ (green) at $r=219$. The gray area represents a non-attracting periodic traveling roll set. In (b) and (e), the power spectra for the time series of $Re(\omega_{101})$ are shown for $r=219$ and $r=220$, respectively. (d) Periodic traveling roll attractor $A^{PTR}$ for $r=220$. Figures (c) and (f) are enlargements of the box regions in (a) and (d), respectively.}
 \label{Fig3}
\end{figure}

Figure \ref{Fig3} contains a summary of what happens in the transition from periodic rolls (PR) to periodic traveling rolls near $r=220$. In Fig. \ref{Fig3}(a), the PR attractors $A^{PR}$ are shown in the phase space projection formed by the real and imaginary parts of $\omega_{101}$ for $r=219$. Several symmetric attractors coexist, five of which are shown in different colors, $A_1^{PR}$ (red), $A_2^{PR}$ (green), $A_a^{PR}$ (blue), $A_b^{PR}$ (black) and $A_c^{PR}$ (magenta). They all have a power spectrum like the one shown in Fig. \ref{Fig3}(b) and an ``eight-like" shape as in Fig. \ref{Fig3}(c), which is an enlargement of the small box indicated in \ref{Fig3}(a). The gray area in \ref{Fig3}(a) represents the long initial transient before the trajectory of a random initial condition converges to attractor $A_b^{PR}$. This transient is constituted by a non-attracting periodic traveling roll. Figure \ref{Fig3}(d) shows the PTR attractor $A^{PTR}$ (purple) at $r=220$; the similarity of this attractor with the gray transients of \ref{Fig3}(a) confirms the change in stability of the periodic traveling rolls. The Power Spectral Density (PSD) of $A^{PTR}$ is shown in \ref{Fig3}(e), where the discrete peaks at a basic frequency and its harmonics reveal that $A^{PTR}$ is formed by translating a periodic attractor in the phase space. Figure \ref{Fig3}(f) shows an enlargement of the small box in \ref{Fig3}(d). The structure of this attractor resembles a typical quasi-periodic attractor, however, its PSD does not present a second incommensurable frequency characteristic of a quasi-periodic attractor \cite{Stoica2005, Olga2015, Francis2020b}.

\begin{figure} [htp!]
 \centering
 \includegraphics[width=0.6\textwidth]{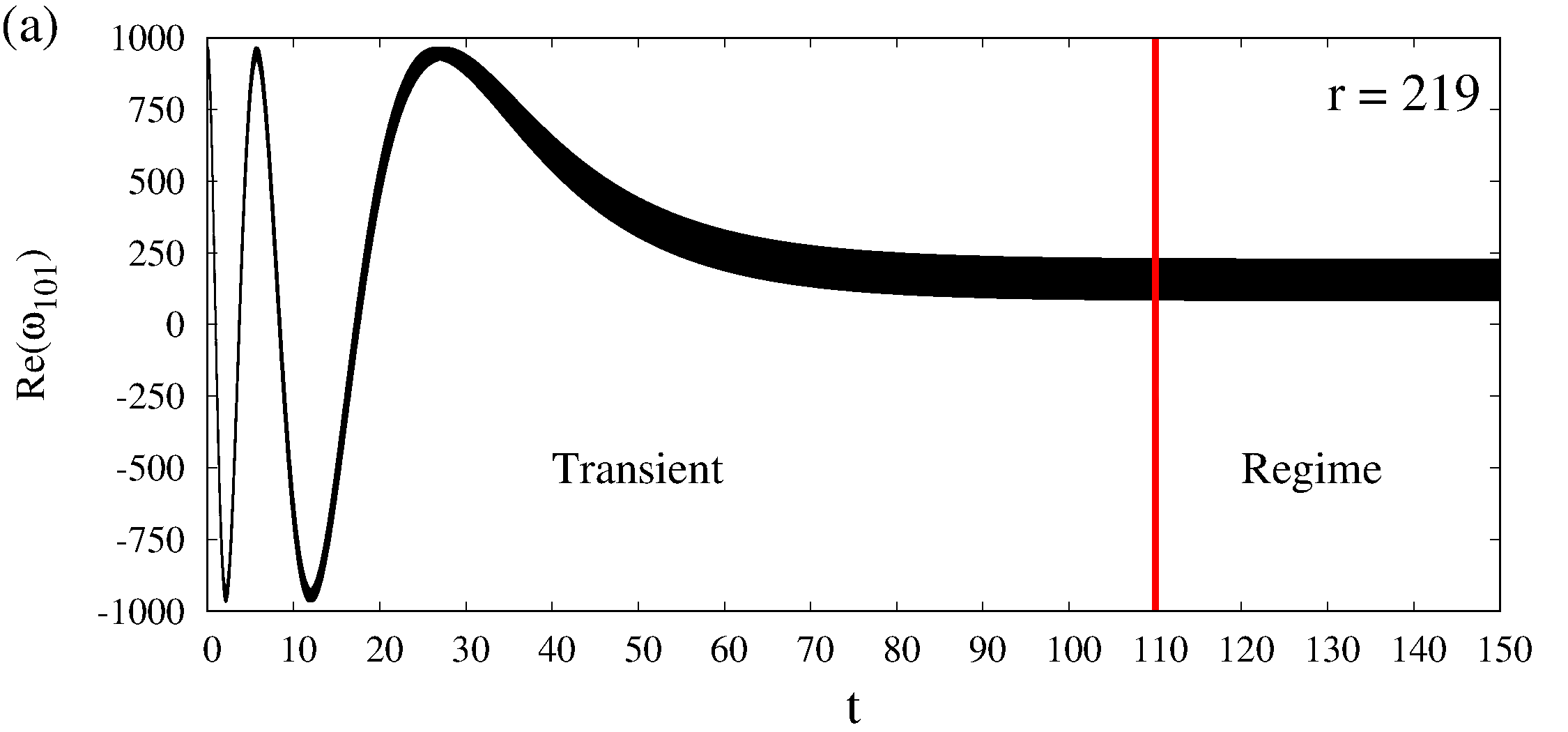}
 \includegraphics[width=0.6\textwidth]{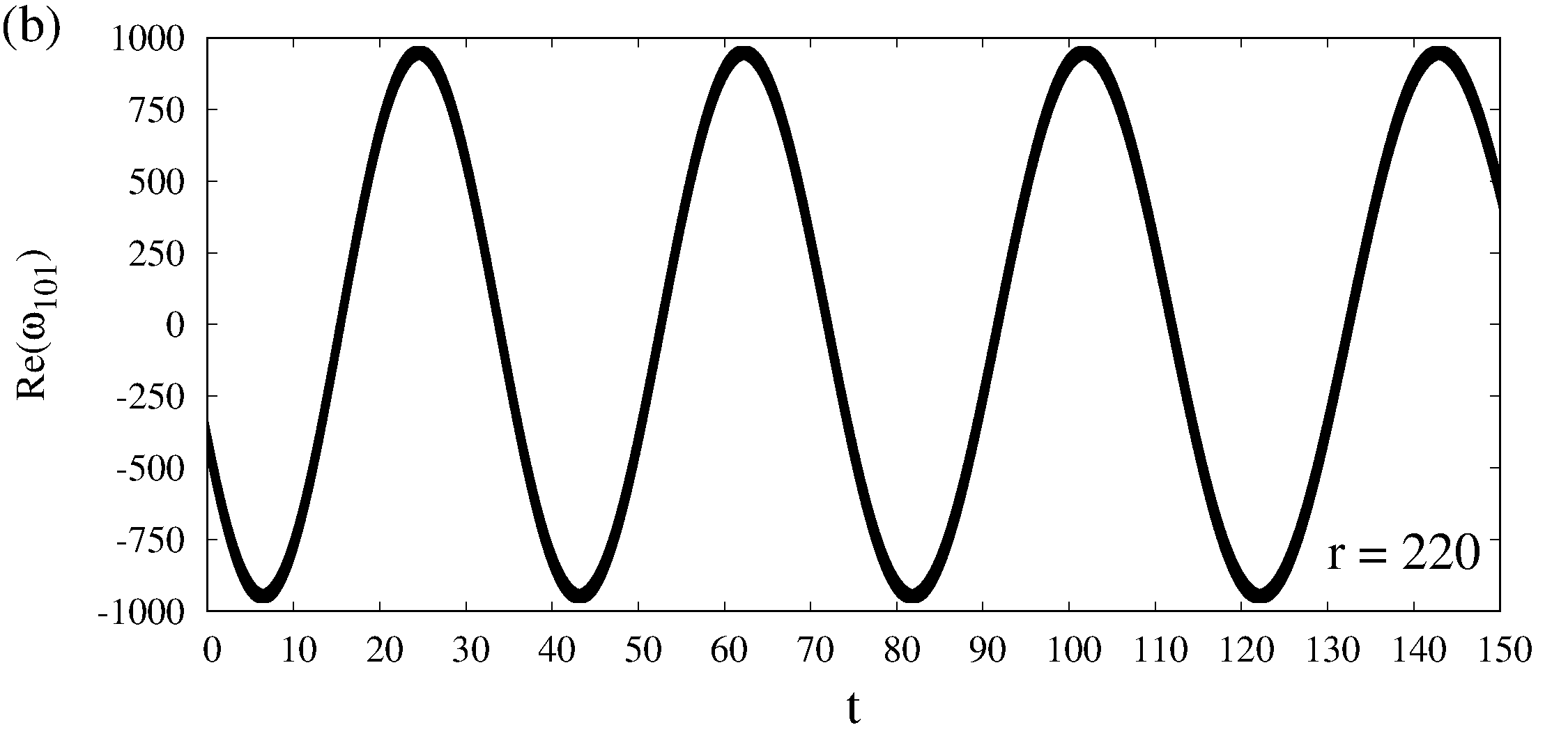}
 \caption{Time series of the real part of the highest energy mode of vorticity $\omega_{101}$ for $r=219$ and $r=220$, in (a) and (b), respectively. In (a) is shown the long initial transient before convergence to the periodic attractor $A_b^{PR}$. In (b), the time series corresponds to the periodic traveling roll attractor $A^{PTR}$. This initial transient in (a) is constituted by a non-attracting periodic traveling roll.}
 \label{Fig4}
\end{figure}

Figure \ref{Fig4} illustrates the time series of the highest energy mode of vorticity, $Re(\omega_{101})$, as a function of time $t$ for two different values of $r$, $r=219$ in Figure \ref{Fig4}(a) and $r=220$ in Figure \ref{Fig4}(b). In Figure \ref{Fig4}(a), the longitudinal red line separates the initial transient from the regime. %It is commonly observed in the literature that long transient times are often caused by the presence of a chaotic saddle in the phase space. Initially, trajectories generated from random initial conditions orbit around the vicinity of the chaotic saddle before eventually converging to an attractor within the phase space. Consequently, during the initial long transient, the trajectory exhibits a structural resemblance to the chaotic saddle itself, which in turn resembles the chaotic attractor that exists at $r=220$.

During the transition from $A^{PR}$ to $A^{PTR}$ ($r=219$ to $r=220$), we observe a similar behavior in the long initial transient of Figure \ref{Fig4}(a) as compared to the time series of the attractor $A^{PTR}$ shown in Figure \ref{Fig4}(b). This similarity becomes more apparent when examining the attractors in the phase space presented in Figure \ref{Fig3}. The gray area in Figure \ref{Fig3}(a), as mentioned earlier, represents the transient observed in Figure \ref{Fig4}(a). The attractor $A^{PTR}$ for $r=220$ bears resemblance to the transient constituted by a non-attracting periodic traveling roll in $r=219$.

\begin{figure} [htp!]
 \centering
  \includegraphics[width=0.3\textwidth]{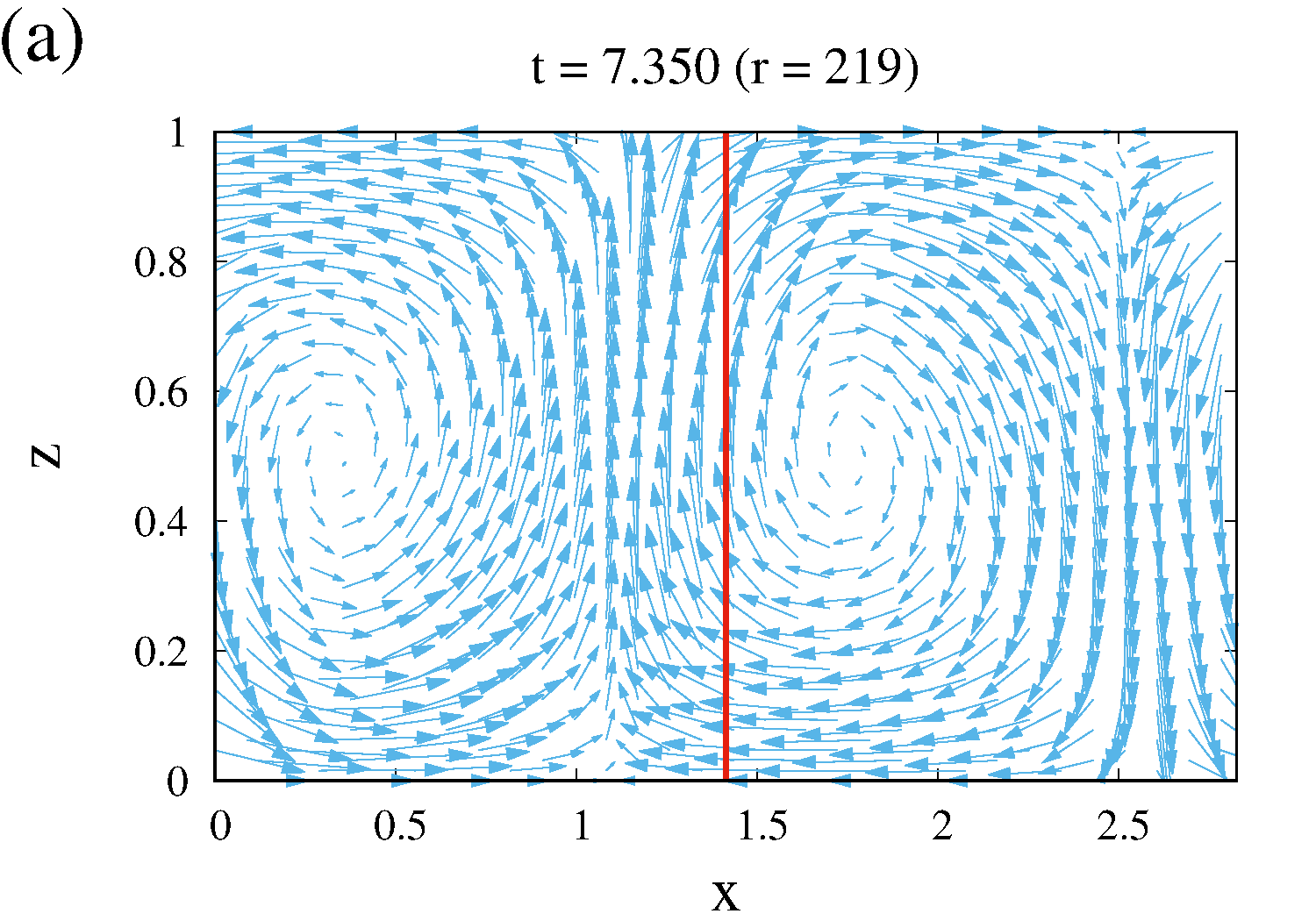}
  \includegraphics[width=0.3\textwidth]{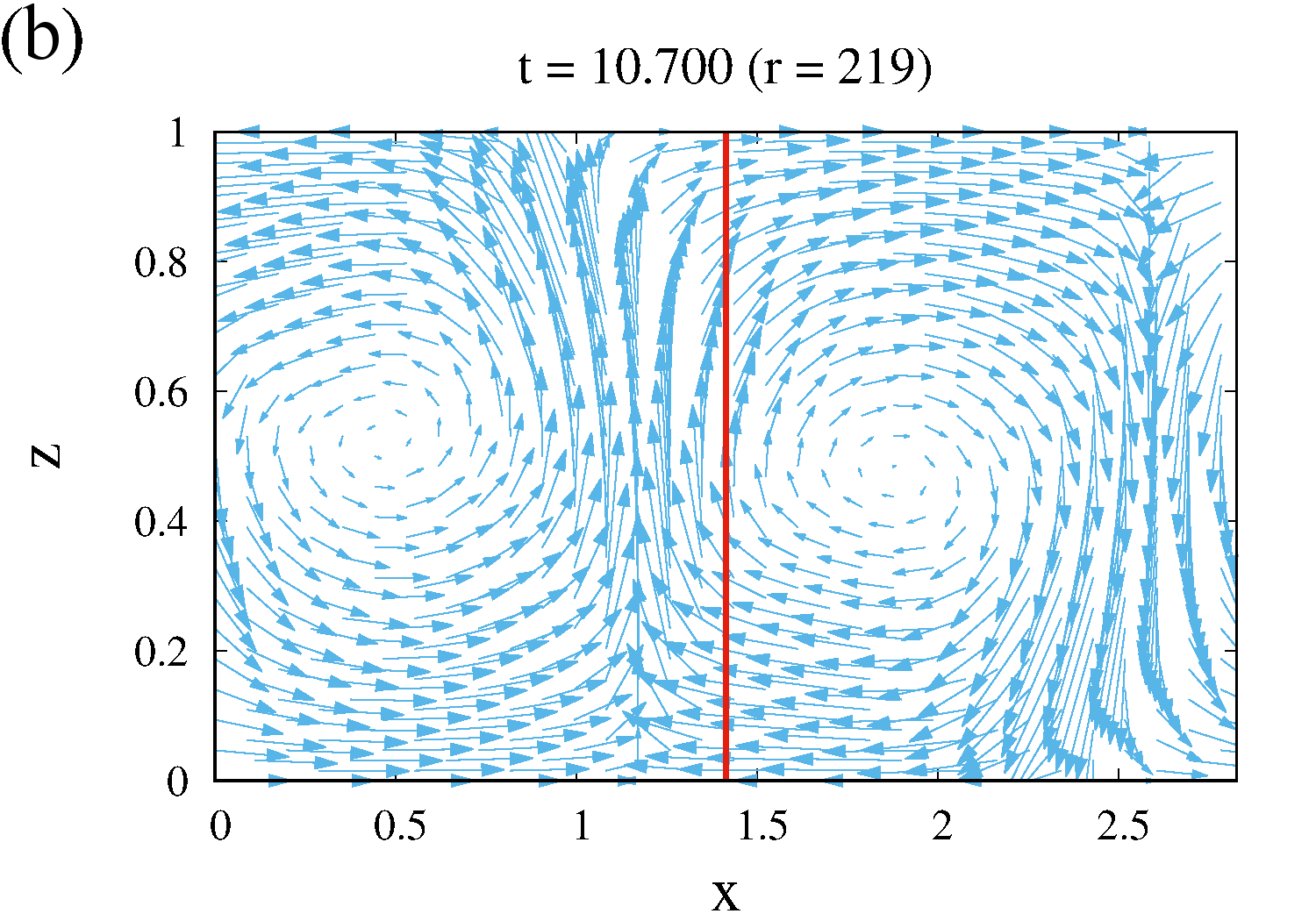}
  \includegraphics[width=0.3\textwidth]{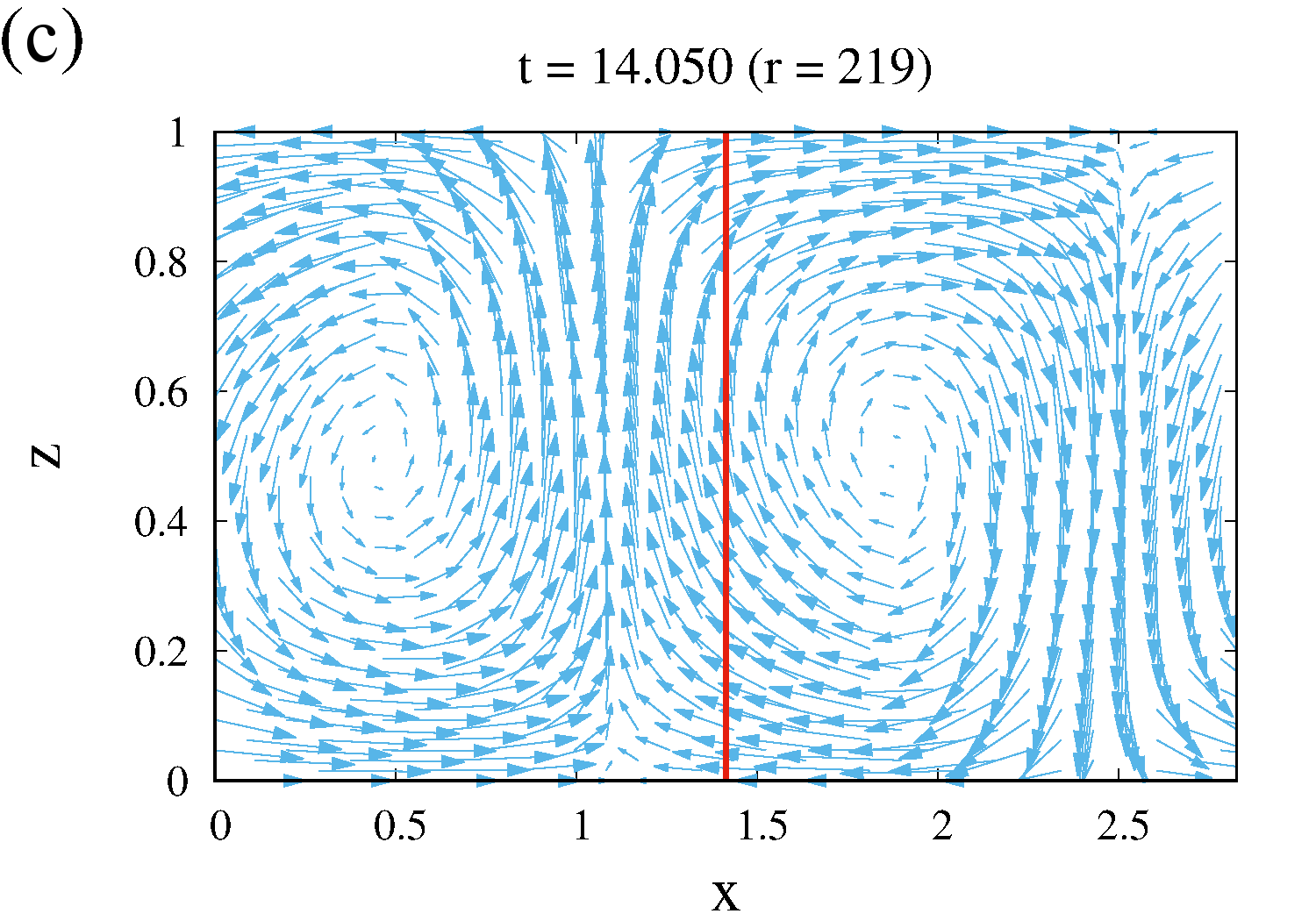}
  \includegraphics[width=0.3\textwidth]{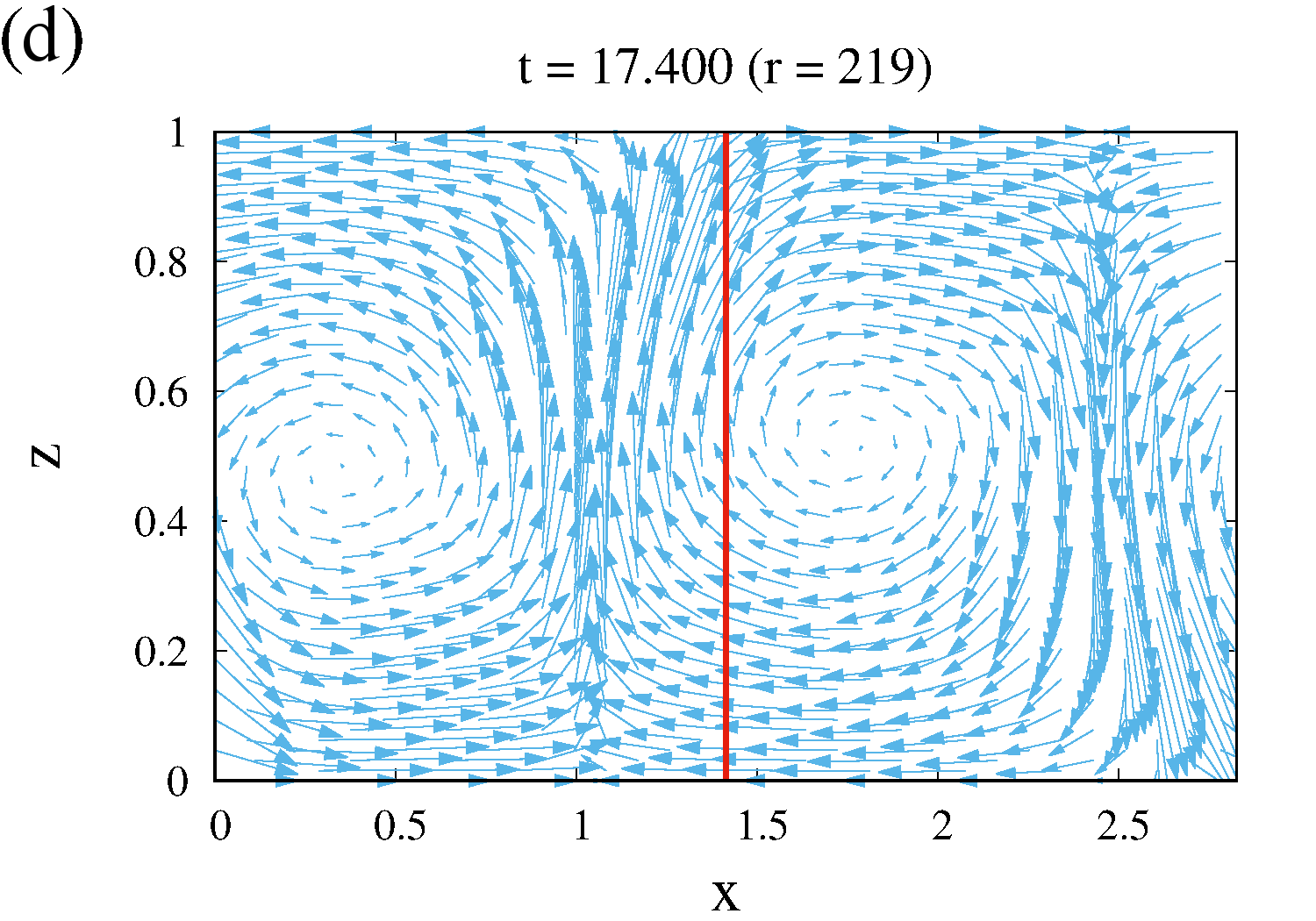}
  \includegraphics[width=0.3\textwidth]{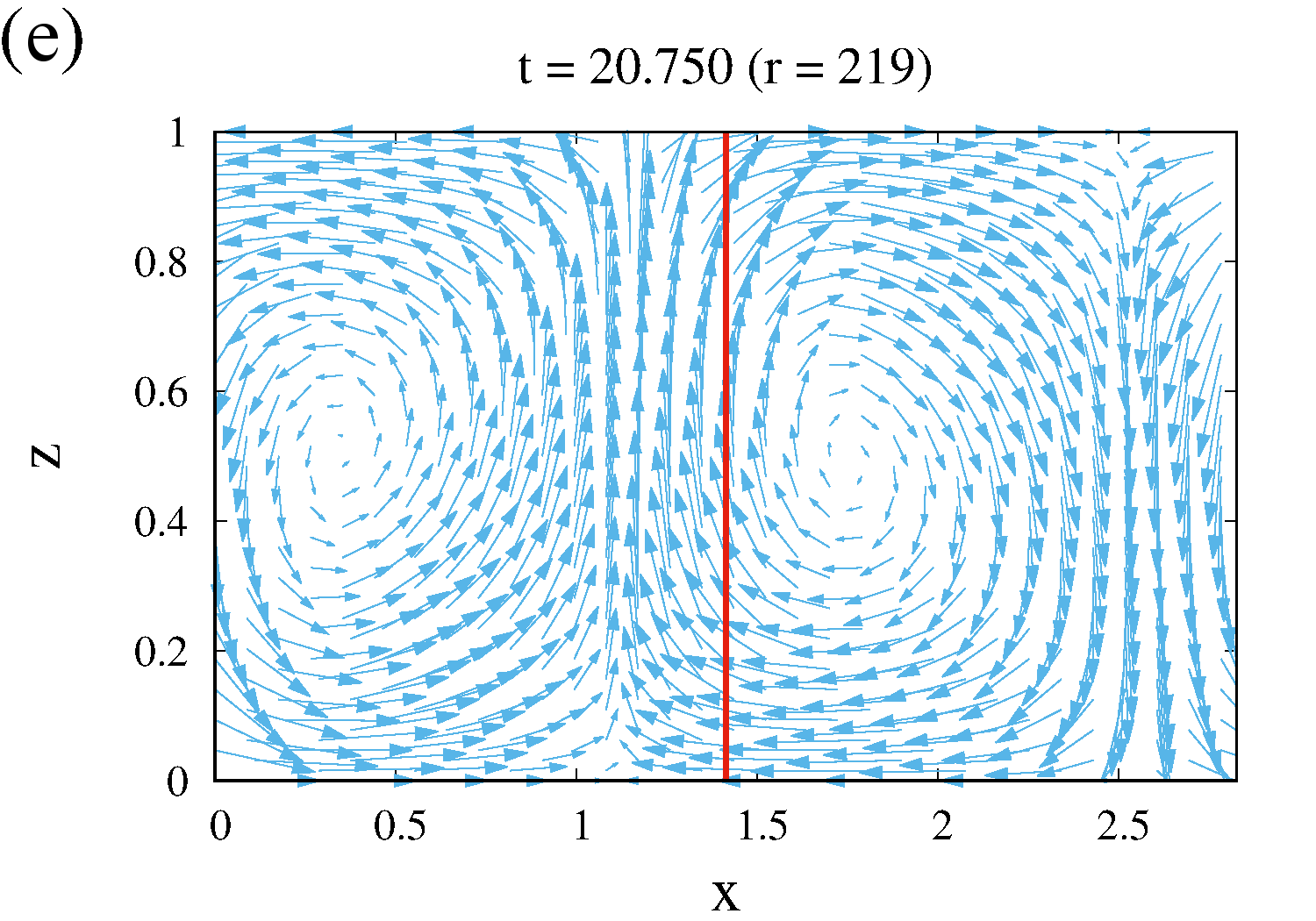}
  \includegraphics[width=0.3\textwidth]{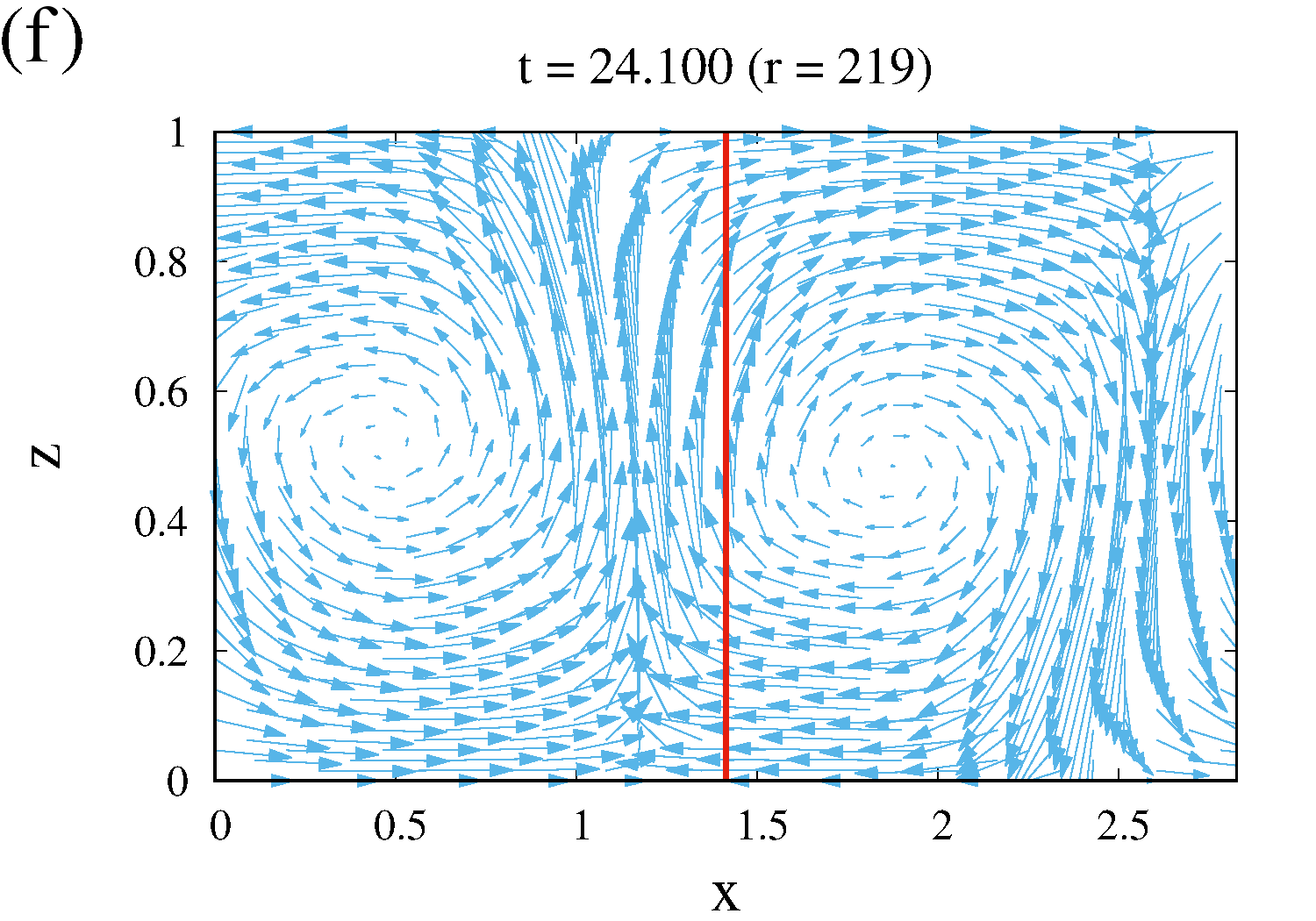}
  \includegraphics[width=0.3\textwidth]{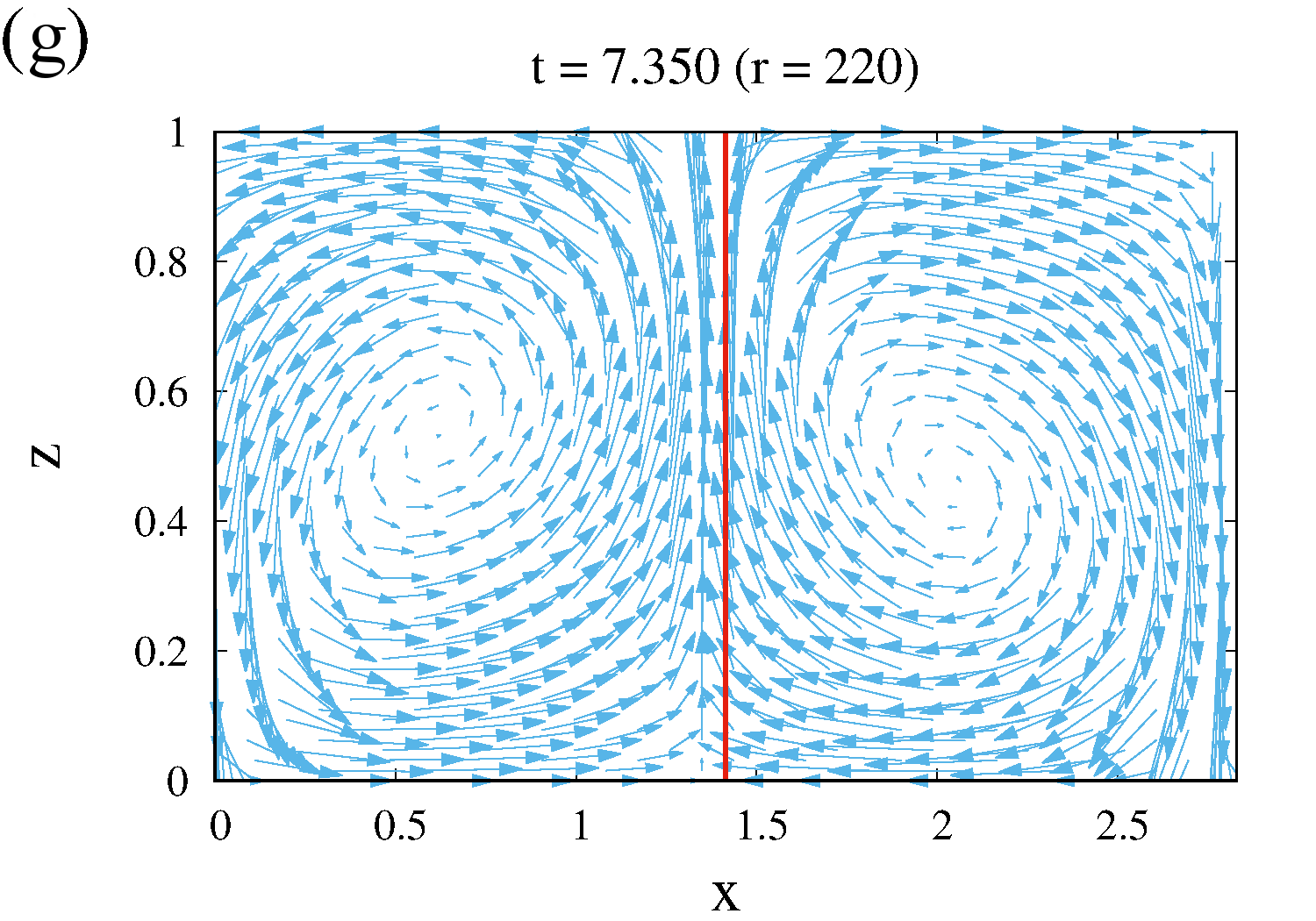}
  \includegraphics[width=0.3\textwidth]{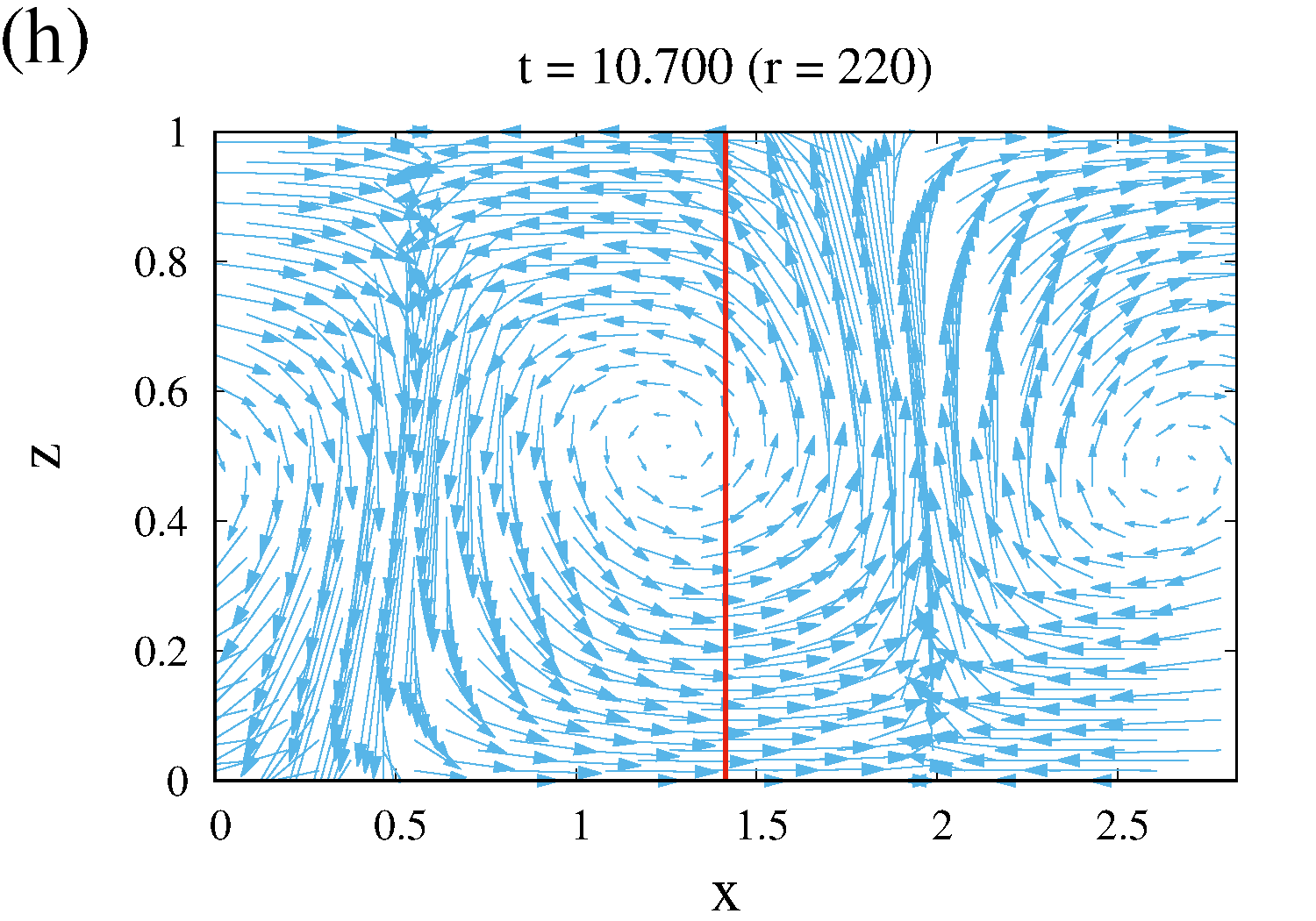}
  \includegraphics[width=0.3\textwidth]{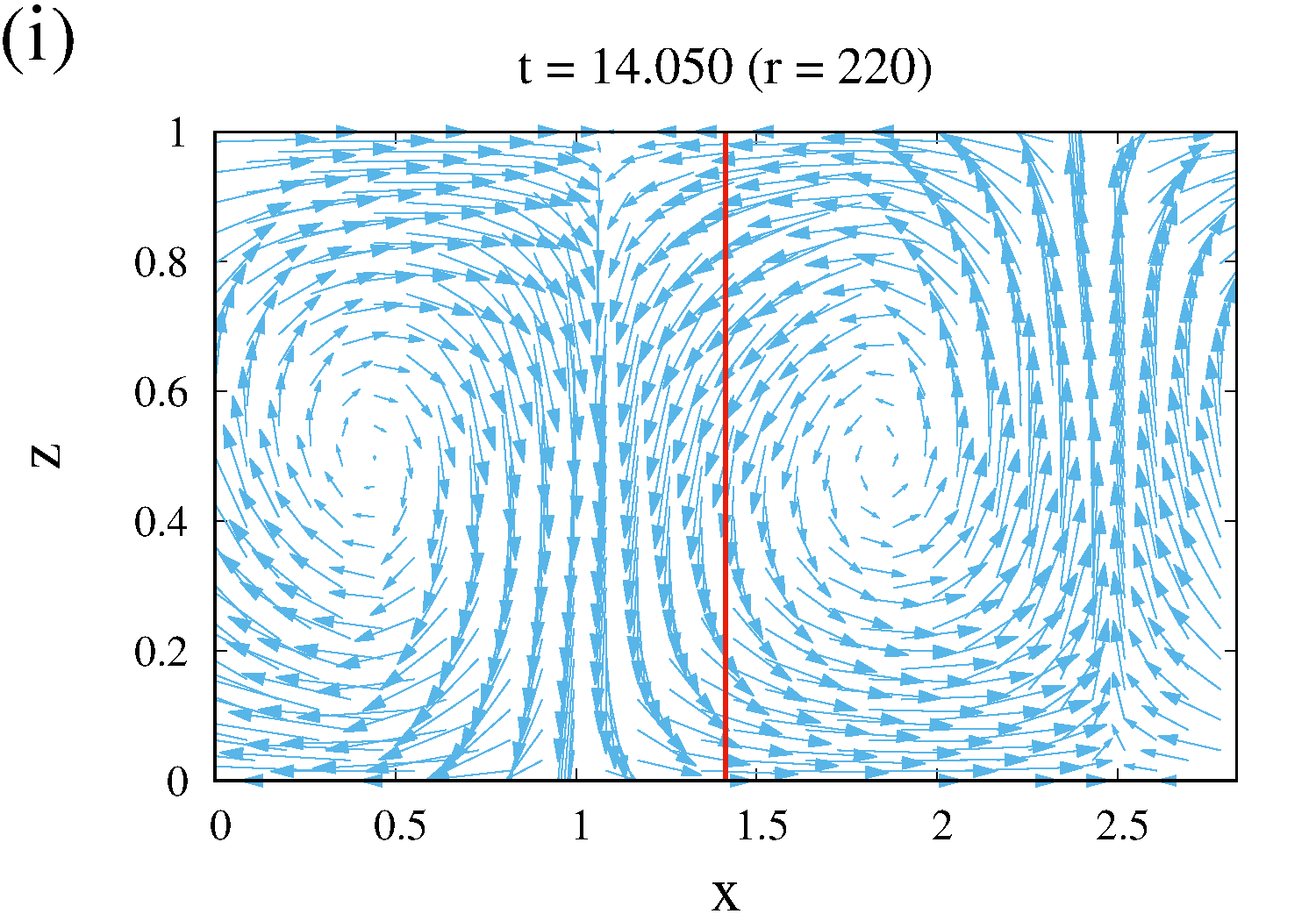}
  \includegraphics[width=0.3\textwidth]{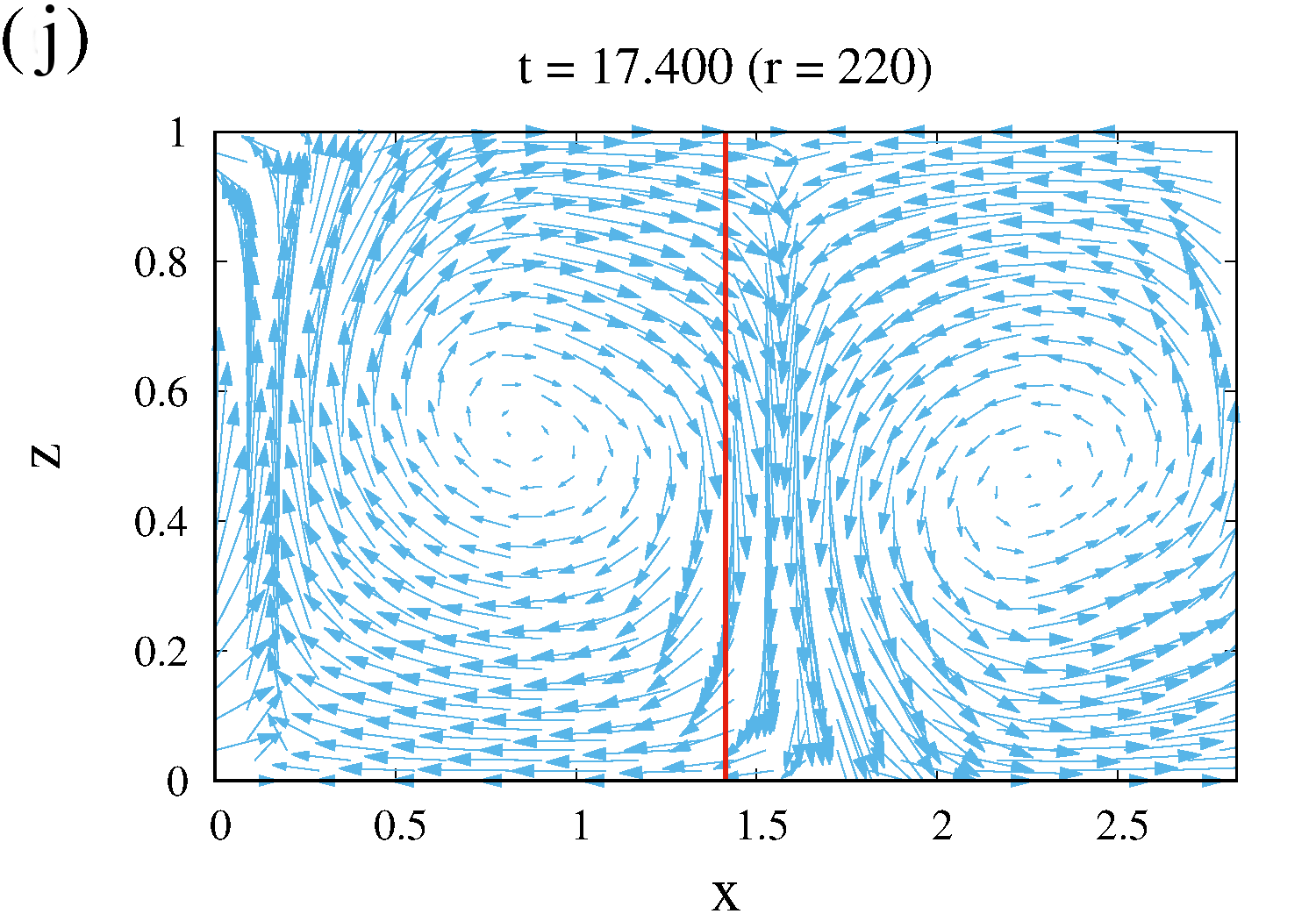}
  \includegraphics[width=0.3\textwidth]{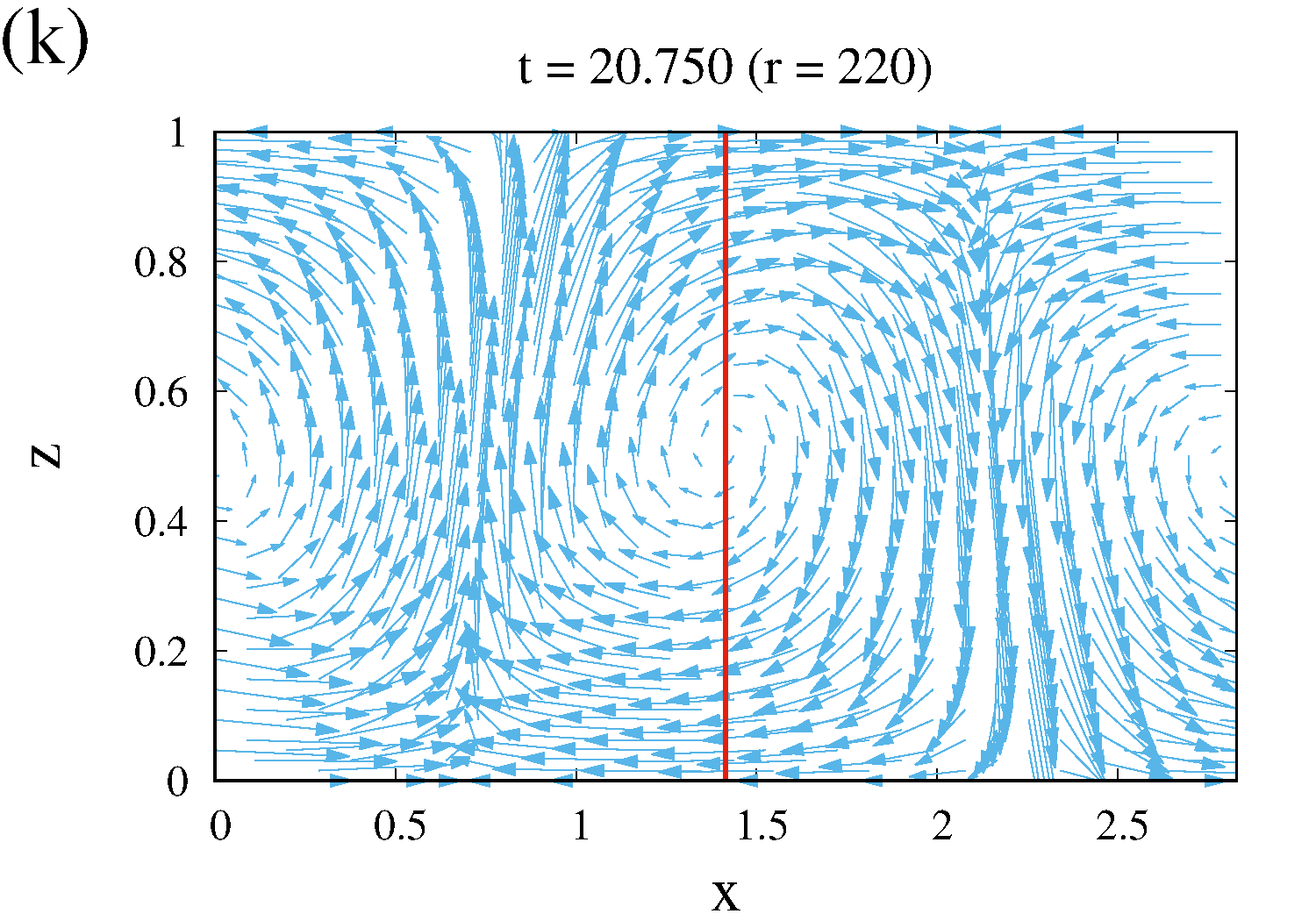}
  \includegraphics[width=0.3\textwidth]{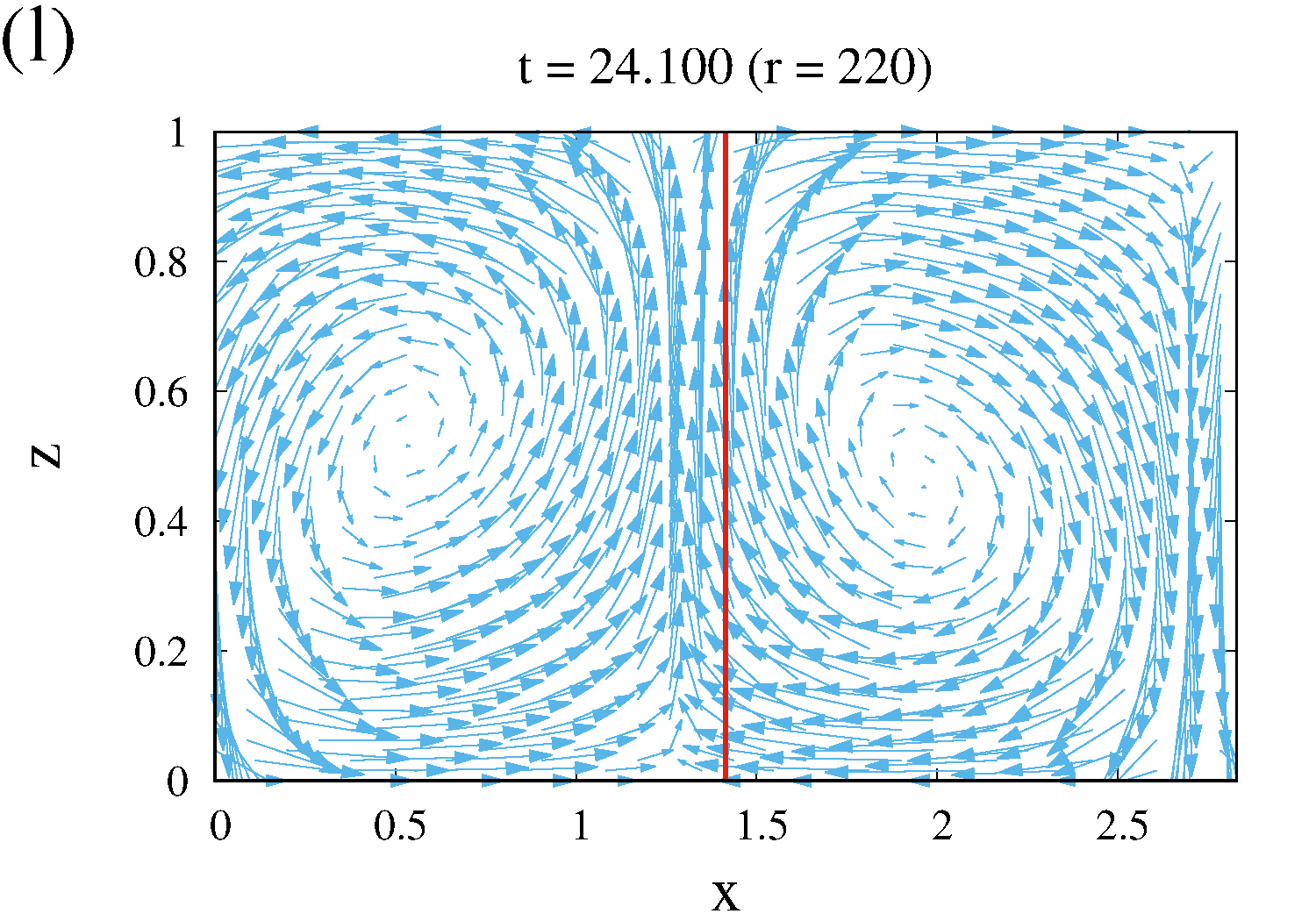}
 \caption{Velocity fields in different instants for $r=219$ (in (a), (b), (c), (d), (e) and (f)) and $r=220$ (in (g), (h), (i), (j), (k) and (l)).}
\label{Fig5}
\end{figure}

The difference between $A^{PR}$ and $A^{PTR}$ in the real space can be appreciated in Figure \ref{Fig5}, where the velocity field is plotted in the $xz$ plane for different values of time and $r$. In Figs. \ref{Fig5}(a) to \ref{Fig5}(f), a PR attractor at $r=219$ is depicted for six consecutive instants. The vertical red lines indicate the middle of the spatial domain and serve as a guide to the eye, revealing that the two large velocity field rolls oscillate periodically in time, but their interface remains near the middle of the box. On the other hand, for periodic traveling rolls at $r=220$, Figs. \ref{Fig5}(g)--\ref{Fig5}(l) show that, besides the periodical oscillations of the rolls, the same are advected from left to right at a constant velocity.

\begin{figure} [htp!]
 \centering
   \includegraphics[width=0.9\textwidth]{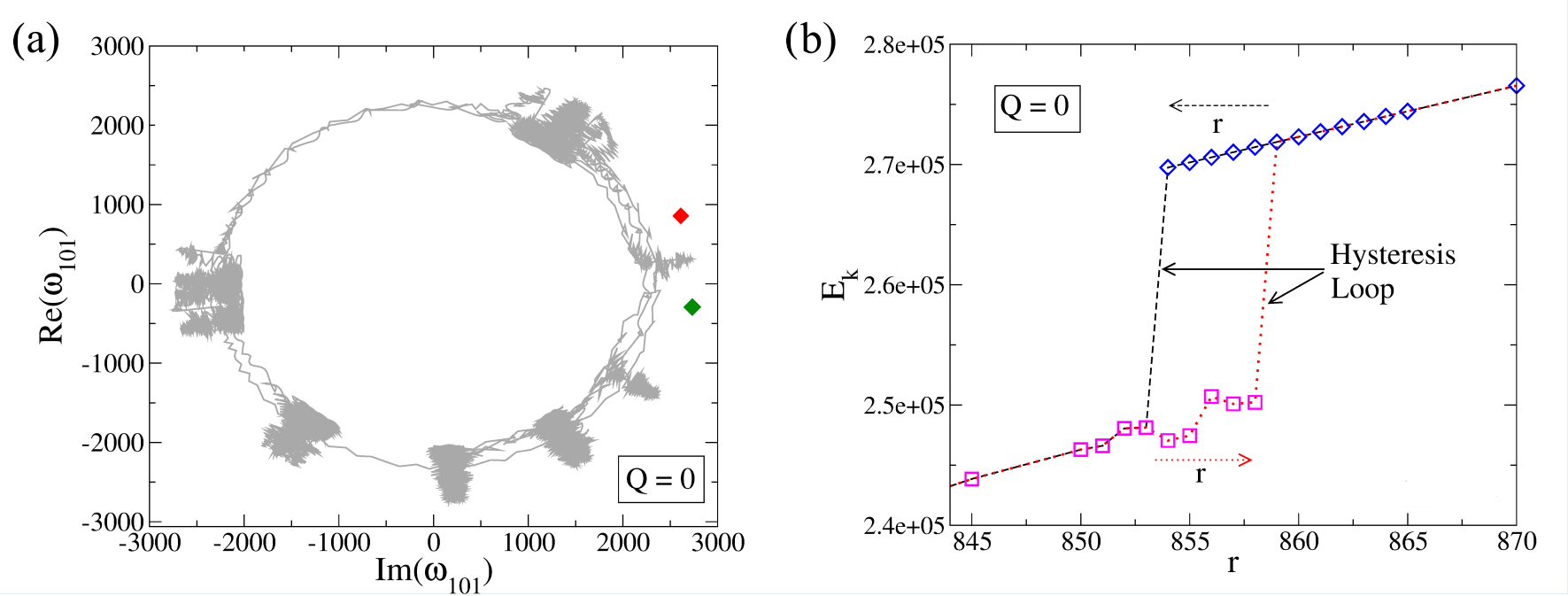}
 \caption{(a) Multistability in the $Im(\omega_{101}) \times Re(\omega_{101})$ phase space projection at $r=857$. The red and green diamonds represent the $A_1$ and $A_2$ fixed point projections, respectively, and the gray lines represent a chaotic traveling rolls attractor. (b) Hysteresis diagram showing the time-averaged kinetic energy $E_k$ of the attractors as a function of $r$. The magenta squares represent the chaotic traveling rolls attractors, and the blue diamonds represent the fixed point attractors.}
\label{Fig6}
\end{figure}

As mentioned before, a chaotic attractor is formed at $r\approx 725$. Actually, there are several symmetric chaotic attractors in the phase space; at $r\approx 840$, these chaotic attractors simultaneously collide with their basin boundaries and undergo a merging crisis, leading to an enlarged attractor. (Such crisis-induced transitions, where attractors collide with invariant saddle sets leading to suddenly enlarged chaotic or hyperchaotic attractors, are a hallmark of complex convective flows \cite{Roman2015}). The new attractor presents chaotic oscillations and horizontal translations of the convective rolls in the physical domain. For that reason, they are called chaotic traveling rolls (CTR). Unlike the periodic traveling rolls, CTR do not move uniformly from left to right, but may suffer intermittent interruptions in the translation motion and even switch directions during their evolution. Such intermittent behavior is a fundamental characteristic of convective regimes during the transition to turbulence. For instance, in three-dimensional Rayleigh-B\'enard convection, similar intermittency in convective attractors has been shown to directly drive intermittent dynamo action \cite{Roman2017}. Figure \ref{Fig6}(a) shows a CTR attractor (gray) in the phase space projection formed by the real and imaginary parts of the vorticity mode $\omega_{101}$ at $r=857$.
The plot reveals the union of several ``nodes'' of chaotic oscillations within a circular region of the phase space. These nodes represent the regions previously occupied by some of the chaotic attractors that were united during the merging crisis at $ r=840,$ and the gray lines between them represent intermittent travel between nodes. For longer orbits, more nodes appear in the plot.
Figure \ref{Fig6}(a) also shows that two symmetric fixed point attractors, denoted as $A_1$ (red diamond) and $A_2$ (green diamond), coexist with the CTR attractor.
For $854 < r < 858,$ the system displays this multistability, which is responsible for the hysteresis curve shown in Fig. \ref{Fig6}(b), where the time-averaged kinetic energy $E_k$ is shown as a function of $r$. The magenta squares represent the CTR attractor, and the blue diamonds represent the fixed point attractors. If an initial condition is chosen in the vicinity of the CTR attractor and $r$ is increased in small steps from $r=845$, the solution will always converge to the CTR attractor up to $r=858$, when the attractor loses stability, and trajectories jump to the upper energy branch of fixed point attractors. On the other hand, if $r$ is reduced from $r=858$, a solution started in the upper branch does not return to the lower branch, but stays in the upper branch until $r\approx 854$, when the fixed point attractor disappears, and the solution returns to the lower energy branch.

\subsection{2D Rayleigh-B\'enard Magnetoconvection} \label{mrb}

\subsubsection{Bifurcation diagrams}

\begin{figure} [htp!]
 \centering
 \includegraphics[width=0.4\textwidth]{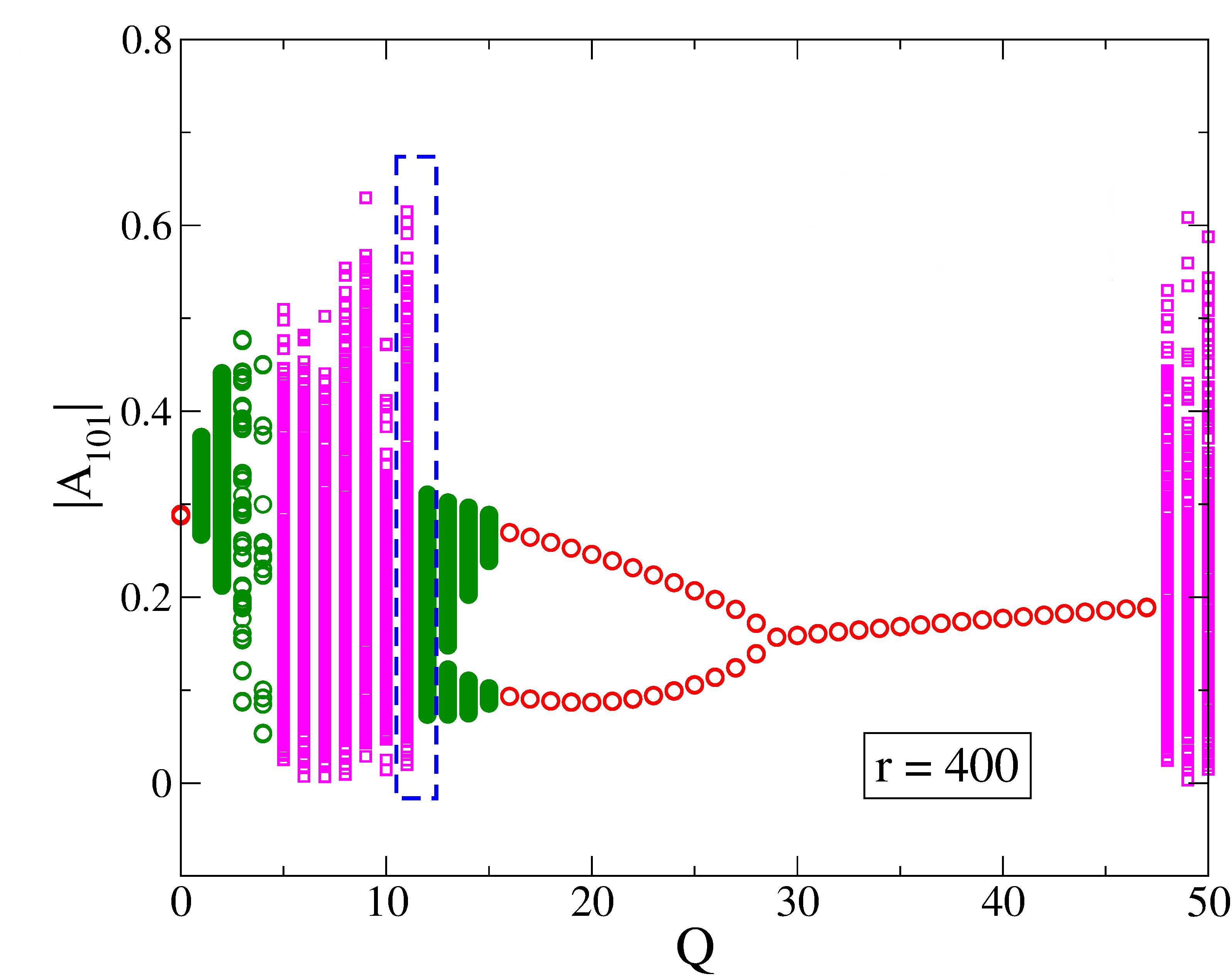}
 \caption{Bifurcation diagram for $|A_{101}|$ as a function of $Q$ for $r=400$. The red circles denote periodic attractors, green circles denote quasiperiodic attractors, and magenta squares denote chaotic attractors.}
\label{Fig7}
\end{figure}

We now move on to the study of convection in the presence of a background magnetic field. %Although the terms ``roll" and ``vortex" appear in the literature with the same meaning, in this work, convection rolls correspond only to the velocity field, while ``vortex'' is used for the magnetic field. We made this distinction only to help describe the observed behaviors. %We first investigate the roll of the magnetic field by fixing $r$ and increasing $Q$ from 0 to 50.
First, we investigate the system's behavior by fixing $r$ and increasing $Q$ from $0$ to $50$. It can be seen from Fig. \ref{Fig2} that for $r=400$ there is a periodic attractor in the purely hydrodynamic system for $Q=0$. The sequence of bifurcations as a function of $Q$ can be seen in Fig. \ref{Fig7}, where the bifurcation diagram of the magnitude of the highest energy Fourier mode for the magnetic vector potential $|A_{101}|$ is plotted as a function of $Q$ for $r=400$. First, there is a traditional Ruelle-Takens route to chaos via quasiperiodicity for $0< Q< 5$. For $Q=0$, the attractor is periodic (red circle), then it becomes quasiperiodic (green circles), then chaotic (magenta squares). At $Q\approx 11.35$, there is a window of quasiperiodic (green) and periodic (red) behavior up to $Q\approx 47$, when the periodic attractor is substituted by the chaotic traveling rolls attractor (magenta). %(see Fig. \ref{Fig7.1}) . 
Hence, the chaotic traveling rolls attractor emerges in the system at the end of the periodic window. The emergence of traveling wave dynamics as a robust feature in convective regimes has also been extensively documented in three-dimensional MHD simulations of rotating convection, where traveling waves interact with magnetic modes to produce complex quasiperiodic and chaotic regimes \cite{Roman2010}.

%We speculate that, from right to left in Fig. \ref{Fig7}, the periodic window is generated by an embedded saddle-node bifurcation (in the Poincar\'e map) at $Q=47$ and is destroyed by an interior crisis-like event involving the collision of a quasiperiodic attractor and a chaotic saddle at $Q\approx 11.35$, but characterizing those bifurcations is not in the scope of the present work. For now, the important conclusion is that a background magnetic field can take a hydrodynamic periodic regime and drive it to a chaotic traveling rolls attractor.
Based on the observed topology of the bifurcation diagram (Fig.~\ref{Fig7}), the periodic window exhibits features consistent with emergence via a saddle-node bifurcation at $Q \approx 47$ and termination via an interior crisis-like event at $Q \approx 11.35$. While a rigorous characterization of these global bifurcations using Poincaré maps falls outside the scope of this work, this sequence is well documented as a route to chaos in extended dissipative systems. Similar sequences of global bifurcations, including precisely characterized interior and boundary crises involving chaotic saddles, have been shown to drive the transition to hyperchaos even in purely hydrodynamic three-dimensional Rayleigh-Bénard convection \cite{Roman2015}. Interior crises, in particular, play a key role in structuring the complex dynamics of magnetohydrodynamic systems. In three-dimensional convective dynamos, an interior crisis, where a quasiperiodic attractor collides with a chaotic saddle, has been shown to form an enlarged chaotic attractor characterized by intermittency, which ultimately dictates the critical magnetic Prandtl number for field generation \cite{Roman2017}. In the present study, the critical finding is that an imposed magnetic field robustly drives a purely hydrodynamic periodic regime into a chaotic traveling-roll state. 

%\begin{figure} [htp!]
% \centering
% \includegraphics[width=0.8\textwidth]{New_figures_gabriel/Fig12-fs-cs.png}
% \caption{Velocity field evolution in a chaotic transient regime for $r=400$ and $Q=11.35$. In background of space physics is ploted of are electric current density $J$.}
%\label{Fig7.1}
%\end{figure} 

\begin{figure} [htp!]
 \centering
  \includegraphics[width=0.45\textwidth]{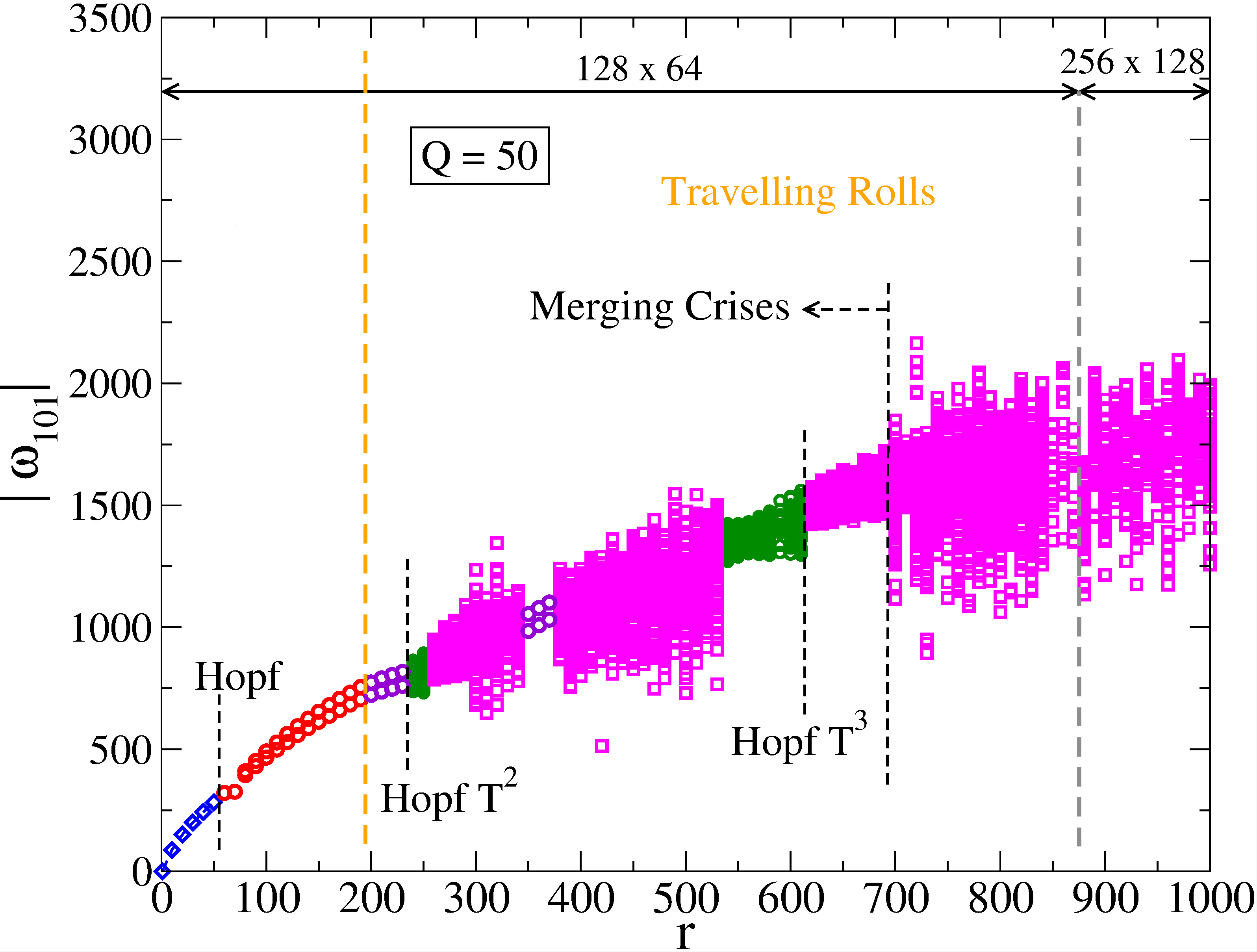}
 \caption{Bifurcation diagram for the magnitude of $\omega_{101}$ as a function of $r$ for $Q=50$. Blue diamonds represent fixed point attractors, red circles denote periodic attractors, purple circles denote periodic traveling rolls, green circles denote quasiperiodic traveling rolls, and magenta squares denote chaotic traveling rolls.}
\label{Fig8}
\end{figure}

Next, we fix the Chandrasekhar number in $Q=50$ and vary the reduced Rayleigh number in the interval $0 \leq r \leq 1000$ to see how the bifurcation diagram of Fig. \ref{Fig2} changes with the addition of a magnetic field. The resulting diagram is shown in Fig. \ref{Fig8}. The first two regimes are very similar to the purely hydrodynamic case, with a fixed point (blue diamonds) bifurcating to a periodic attractor (red circles), then to periodic traveling rolls (purple circles). But from then on, all detected attractors in the range $200 < r \leq 1000$ exhibit traveling rolls behavior in physical space. Thus, for $230 < r < 260$ there is a quasiperiodic traveling rolls attractor (green circles); for $r > 260$, the attractors are chaotic traveling rolls for most values of $r$, with a window of periodic traveling rolls for $350 < r < 380$ and a window of quasiperiodic traveling rolls for $540 < r < 620$. We conclude that the magnetic field favors the traveling wave dynamics in the system. For $0<r<870$ the mid numerical resolution ($128\times 64$) was adopted, but for $r>870$ we employed the high resolution ($256\times 128)$. For both resolutions, monitoring of the kinetic and magnetic energy spectra at high Rayleigh numbers confirmed a rapid decay at high wavenumbers. This ensures that the smallest dissipative scales are adequately resolved without energy pile-up at the truncation limits, justifying the use of the mid-resolution for extensive parameter sweeps.

\subsubsection{Roll breaking}

As mentioned above, the convective rolls can present different dynamics in the real space, such as steady rolls (fixed points in the Fourier space), oscillating rolls (periodic, quasiperiodic, or chaotic), and traveling rolls (periodic, quasiperiodic, or chaotic). For $Q = 0$, such oscillating traveling rolls are robust and their spatial structures are not destroyed with time, with two convective rolls being advected while keeping a length scale determined by the height of the fluid layer, as shown in Fig.~\ref{Fig5}. However, in the presence of a magnetic field, these rolls can undergo breaking and merging phenomena, as shown in Fig.~\ref{fig:velocity_field} for $r = 400$ and $Q = 50$, when the system exhibits a chaotic traveling rolls attractor. 
The left column presents the velocity field vectors at different times, with the background color representing the electrical current density ($J$). The middle column shows the instantaneous vorticity deviation (IVD) for the normalized velocity field, $\boldsymbol{v}/|\boldsymbol{v}|$, where ``hot'' colors identify the vortex cores. The maximum finite-time Lyapunov exponents computed for steady velocity fields at each time are plotted in the right column. 

%Unlike in the purely hydrodynamic regime, where only two focus points mark the centers of the two rolls (see Fig.~\ref{Fig5}), in the hydromagnetic regime, we found up to four focus points, which can be clearly identified in the IVD field in Fig.~\ref{fig:velocity_field}: after roll breaking, two additional vortex centers emerge. Moreover, the FTLE field indicates the formation of a saddle point at the interface between the rolls, and it also reveals that roll breaking gives rise to a homoclinic orbit, an “eight-shaped homoclinic orbit”, around the saddle, evidencing the reconnection of stable and unstable manifolds in a manner consistent with the literature\cite{gonchenko13}.

At $t = 1.2305$, the flow exhibits only the two expected convective rolls seen in the hydrodynamic case. High values of FTLE mark the boundary between them. Starting at $t = 1.2310$, the roll cores begin to stretch diagonally. This deformation is clearly captured by the IVD field, which shows the two roll cores elongating and then breaking into four cores at $t=1.2315$. The four rolls are also seen in the velocity vector field (left column), and, even more clearly, in the FTLE field, in the form of two eight-shaped homoclinic connections \cite{gonchenko13}. In this scenario, there is a saddle point at the center of each ``eight'' structure, and its stable and unstable manifolds connect, forming the two lobes of the ``eight'' structure. Within each lobe, there is a focus point, which in our case is marked by the local maxima (red spots) in the IVD field. Thus, the lobes represent the boundaries of newly created vortices or rolls. These new rolls continue to evolve, and one of the lobes in each ``eight'' starts to shrink. At about $t=1.2332$, two of the four focus points within the rolls simultaneously collide with the corresponding saddle points, destroying two lobes, thus restoring the initial configuration with only two convective rolls. This dynamics is recurrent in time and was only observed in our simulations in the presence of a magnetic field.

%%%%%%% CAMPO DE VELOCIDADE %%%%%%%%%%%%%%%%%%%%%%%%%%%%%%%%%%%%%%%%%%
\Needspace{10\baselineskip}
\begin{figurehere}
\centering

\includegraphics[width=\textwidth]{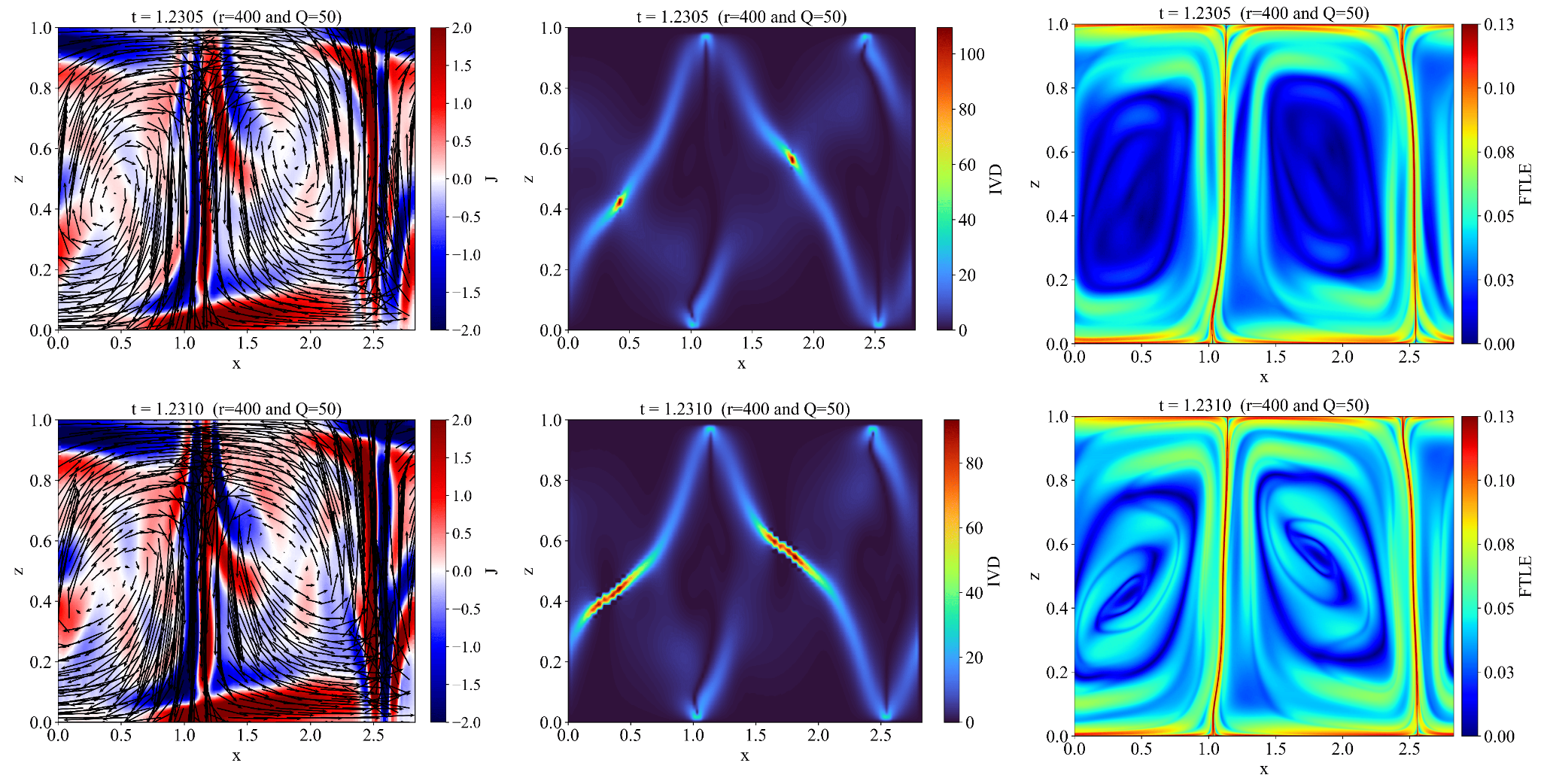}\par\vspace{6pt}
\includegraphics[width=\textwidth]{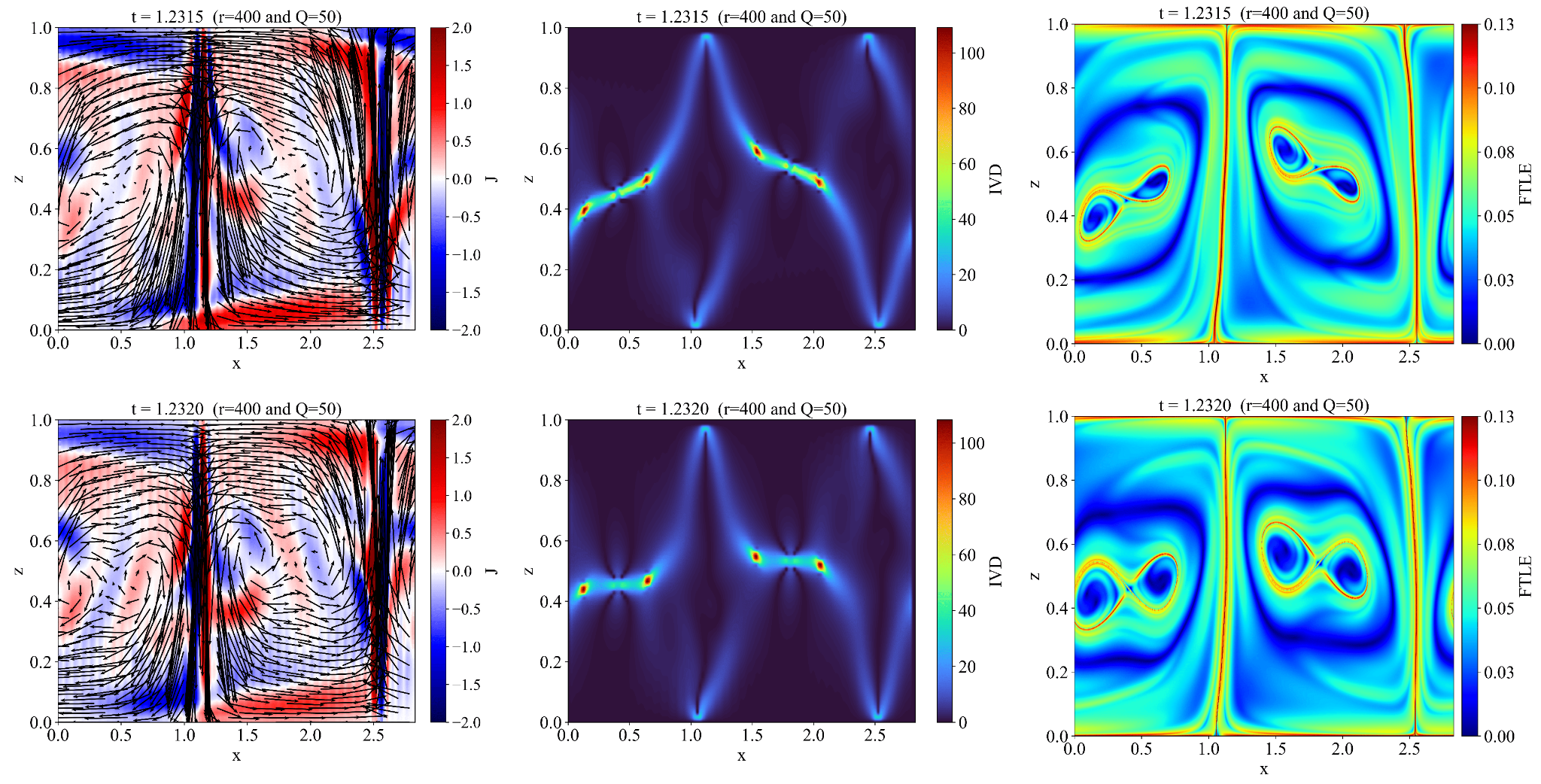}\par\vspace{6pt}
\includegraphics[width=\textwidth]{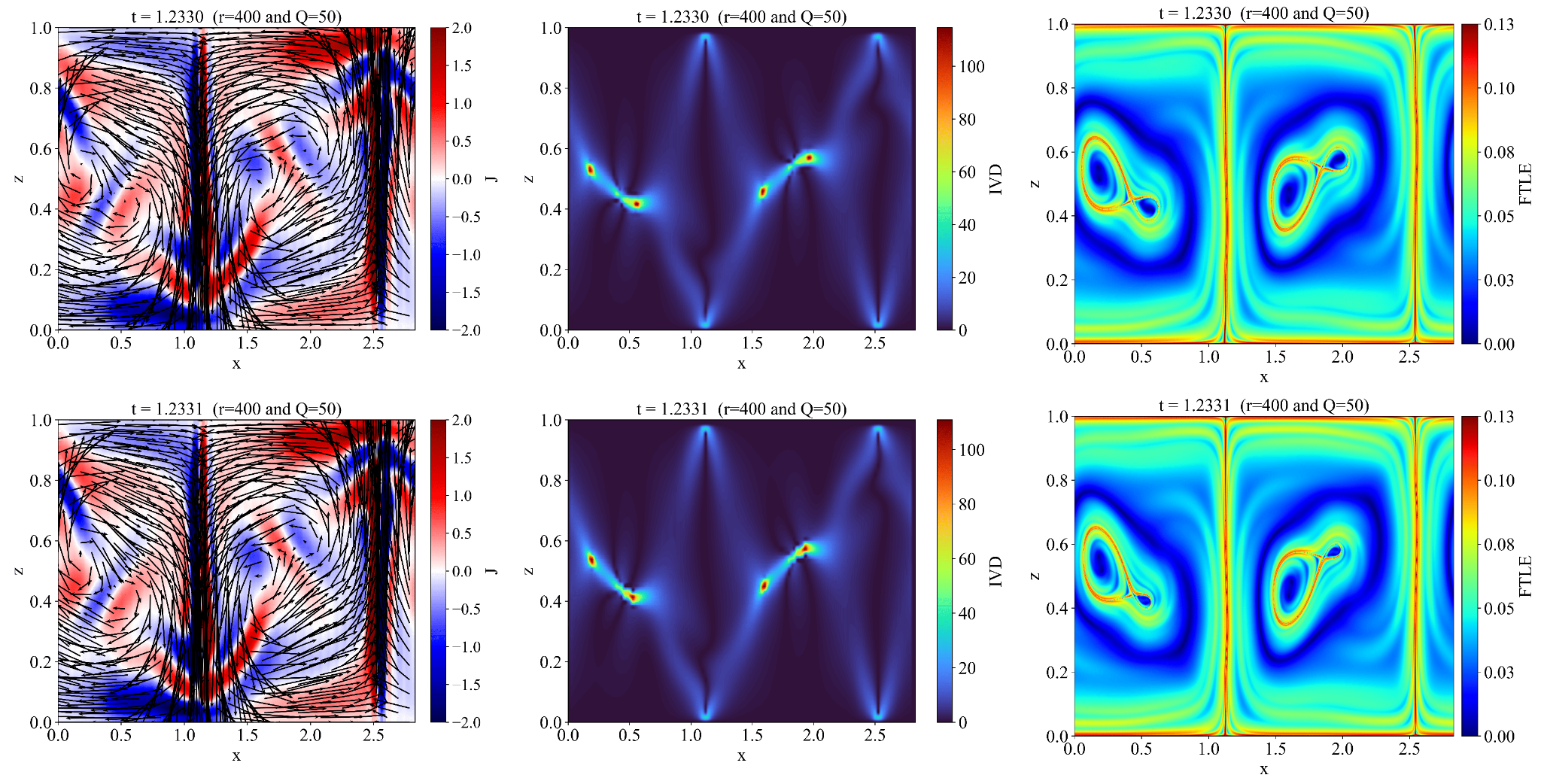}\par\vspace{6pt}
\includegraphics[width=\textwidth]{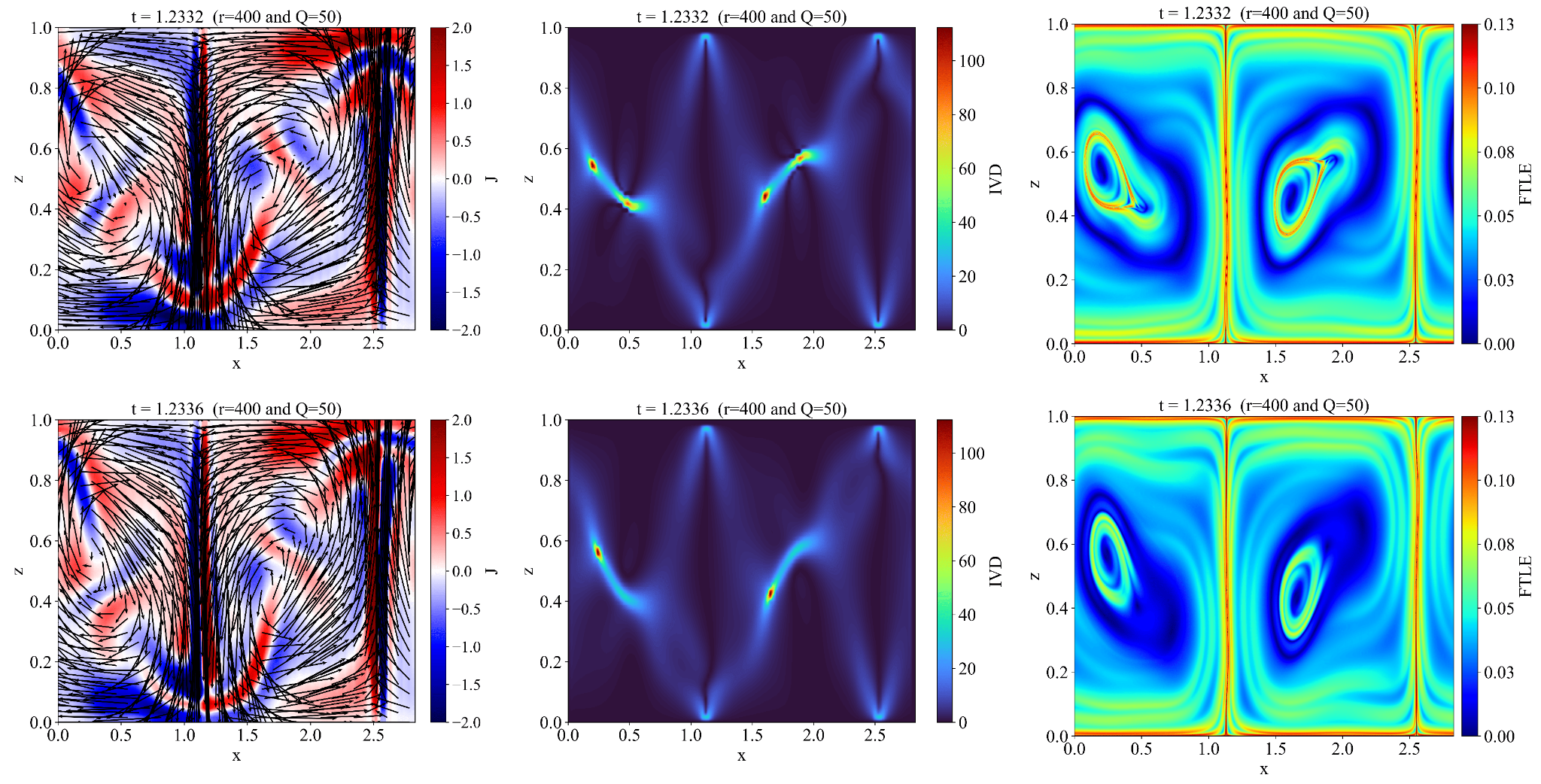}

\caption{Velocity field structures during a vortex break for
$r=400$ and $Q=50$. The left column shows the velocity vectors
with the background colored by the current density; the middle
column shows the IVD field for the normalized velocity field;
the right column shows the maximum FTLE field. Each row
corresponds to a different time.}
\label{fig:velocity_field}
\end{figurehere}
%%%%%%%%%%%%%%%%%%%%%%%%%%%%%%%%%%%%%%%%%%%%%%%%%%%%%%%%%%%%%%%%%%%%%%

Fig.~\ref{fig:magnetic_field} shows the magnetic-field diagnostics at the same time instants as the velocity-field snapshots in Fig.~\ref{fig:velocity_field}. %, allowing us to examine the real-space dynamics at each time. 
 In Fig.~\ref{fig:magnetic_field}, the left panel displays the magnetic vector field, with the current density ($J$) shown as the background colormap; the center panel shows the LCD field computed from the normalized magnetic field, $\boldsymbol{B}/|\boldsymbol{B}|$, with magnetic-field streamlines superimposed; the right panel shows the FTLE field. Overall, the magnetic field exhibits more intricate dynamics than the velocity field, featuring a larger number of magnetic vortices

%%%%%%%% CAMPO MAGNÉTICO %%%%%%%%%%%%%%%%%%%%%%%%%%%%%%%%%%%%%%%%%%%%%%%
\Needspace{10\baselineskip}
\begin{figurehere}
\centering

\includegraphics[width=\textwidth]{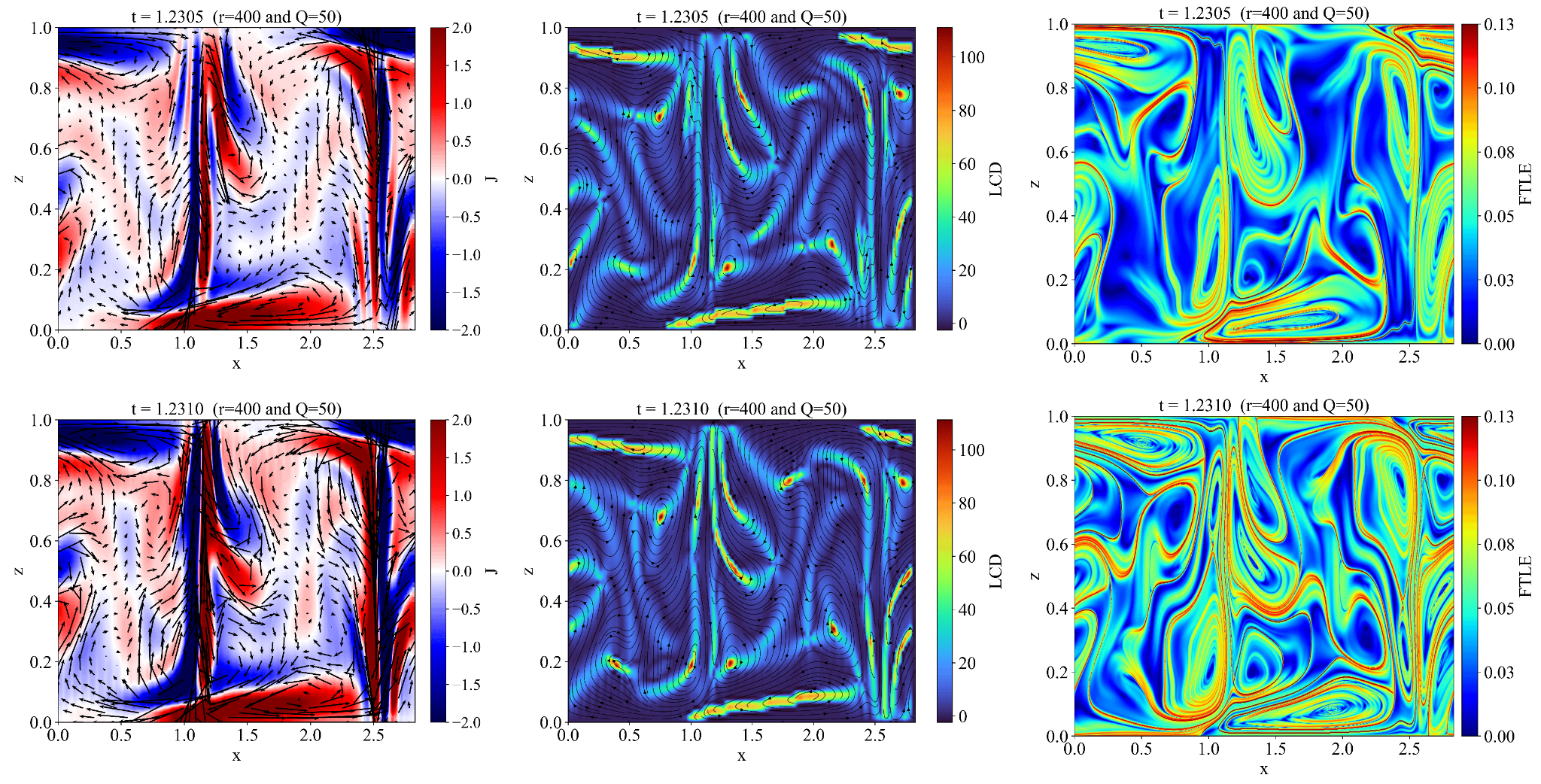}\par\vspace{6pt}
\includegraphics[width=\textwidth]{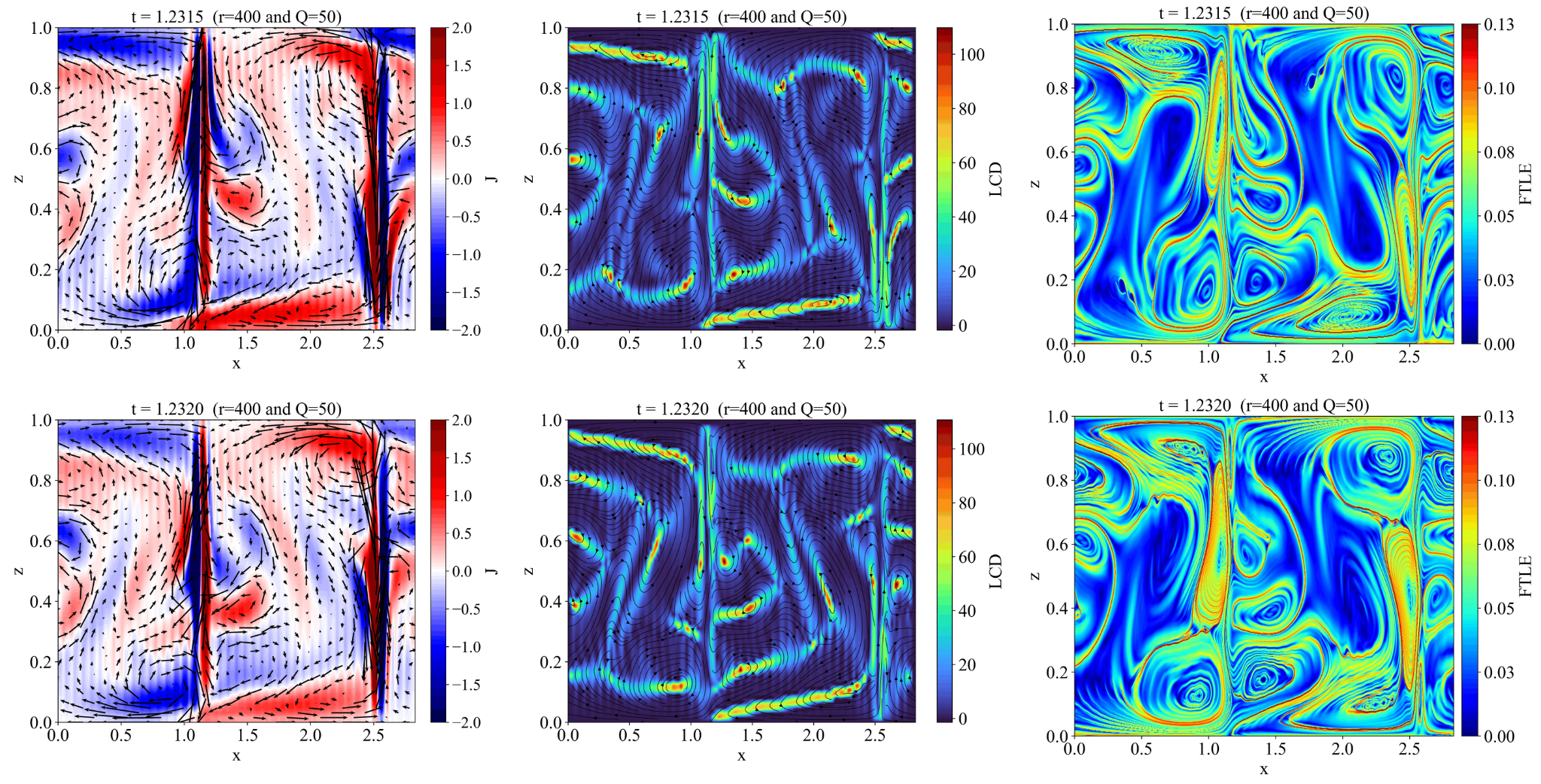}\par\vspace{6pt}
\includegraphics[width=\textwidth]{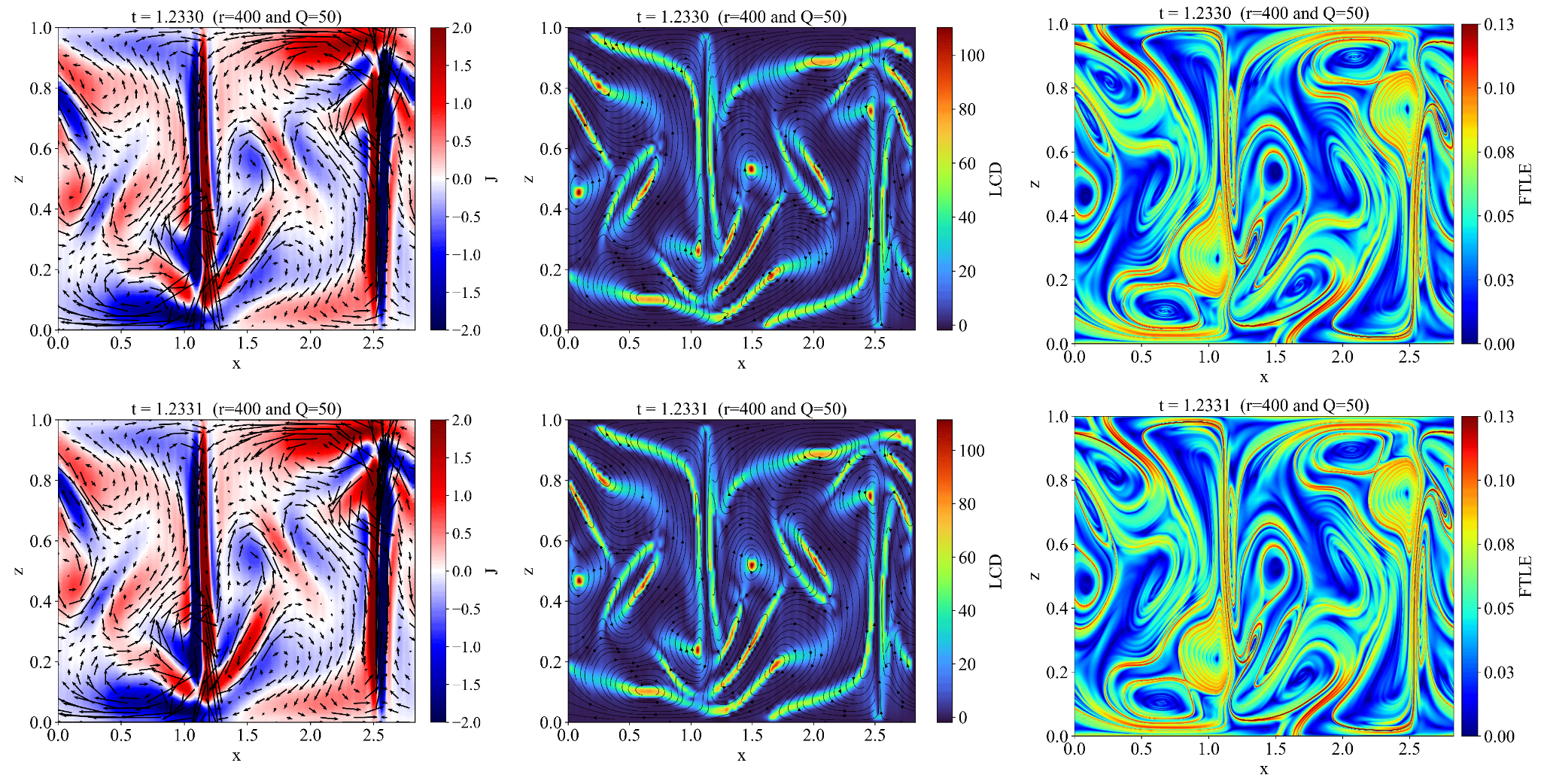}\par\vspace{6pt}
\includegraphics[width=\textwidth]{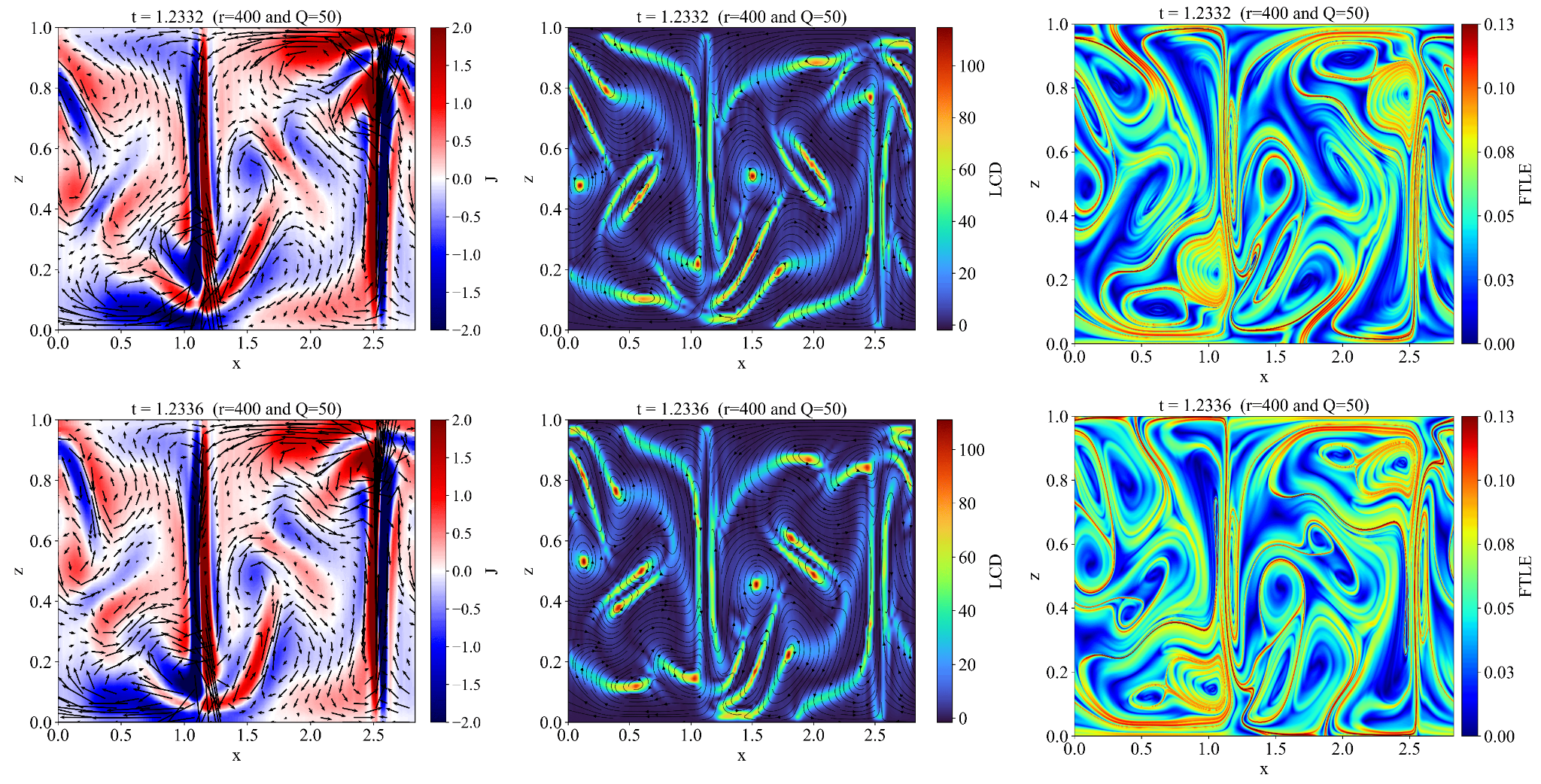}

\caption{Magnetic field structures during the vortex break sequence
of Fig.~\ref{fig:velocity_field}. The left column shows the magnetic
field vectors with the background colored by the current density;
the middle column shows the LCD field superposed by selected
streamlines of the magnetic field; the right column shows the
maximum FTLE field. Each row corresponds to a different time.}
\label{fig:magnetic_field}
\end{figurehere}
%%%%%%%%%%%%%%%%%%%%%%%%%%%%%%%%%%%%%%%%%%%%%%%%%%%%%%%%%%%%%%%%%%%%%%%%%%

\begin{figure} [htp!]
 \centering
  \includegraphics[width=1\textwidth]{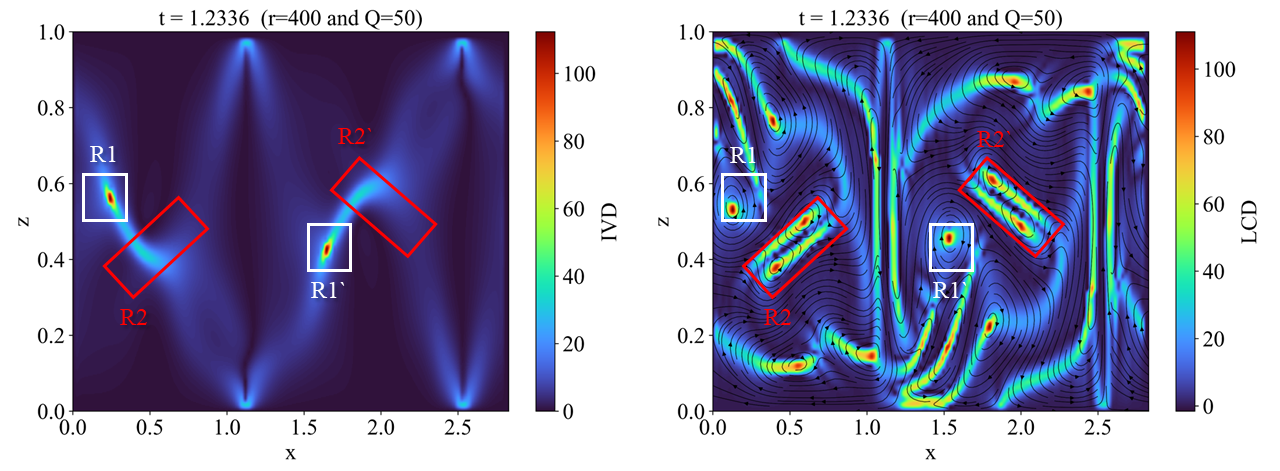}
 \caption{Fields at $t=1.2336$ ($r=400$, $Q=50$). \textbf{Left:} velocity IVD. \textbf{Right:} magnetic LCD with streamlines. Red boxes ($R2$, $R2^{\prime}$) mark the saddle-roll interaction and the co-located magnetic reconnection; white boxes ($R1$, $R1^{\prime}$) indicate non-colliding vortices and their magnetic counterparts.}
\label{fig_ivd_comparison}
\end{figure}

Nevertheless, the magnetic field still exhibits an interesting feature. At approximately $t=1.2336$, when the system returns to its initial configuration, a magnetic reconnection event emerges in the magnetic field, localized in the same region where, in the velocity field, the vortex center collided with the saddle point. Fig.~\ref{fig_ivd_comparison} highlights this correspondence by showing the IVD field for the velocity on the left and the LCD field for the magnetic field on the right. In the $R2$ region (red rectangle), the velocity field's IVD indicates the saddle–roll interaction, and the magnetic field's LCD reveals reconnection in the same spatial location. In addition, vortices in the velocity field that do not undergo collision appear in the same region as magnetic vortices, as emphasized by region $R1$ (white rectangle). Note that, since there are two pairs of vortices, $R1$ and $R2$ correspond to one pair, whereas $R1^{\prime}$ and $R2^{\prime}$ correspond to the other pair.

\begin{figure} [htp!]
 \centering
  \includegraphics[width=1\textwidth]{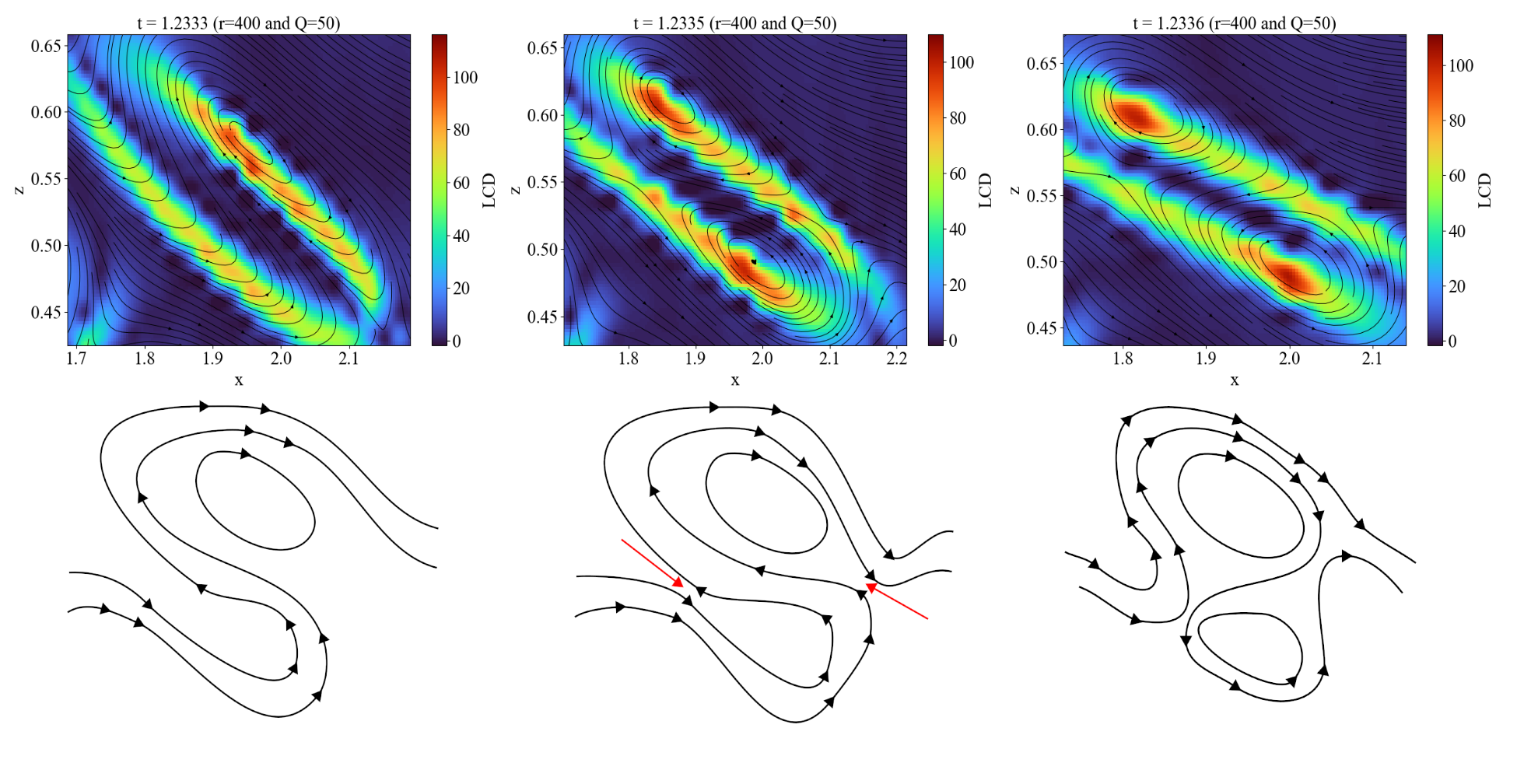}
 \caption{Three consecutive snapshots of the magnetic field lines during a reconnection event; the background is colored by the LCD. 
 The lower panels schematically illustrate the process; red arrows indicate the reconnection sites. }
\label{fig_rx}
\end{figure}

Figure \ref{fig_rx} shows an enlargement of the region near the $R2^{\prime}$ box in Fig. \ref{fig_ivd_comparison}, during the creation of two magnetic vortices. It shows that magnetic vortex duplication is due to a magnetic reconnection event, which results in a new magnetic field topology in which two structures with closed field lines form. The upper panels show the magnetic field lines at three instants, $t=1.2333$ (before reconnection), $t=1.2335$ (at reconnection), and $t=1.2336$ (after reconnection); the background is colored by the LCD. The lower panels show a schematic illustration of the process, with red arrows pointing to the reconnection sites. This figure depicts what happens to the magnetic field precisely at the locus and moment when the focus point collides with the saddle point in the velocity field. It seems that during the approximation of the focus toward the saddle, the velocity field drags the magnetic field lines, squashing the magnetic structure and causing a reconnection in points where magnetic lines with opposite directions collide.

\section{Conclusions}
\label{sec conclusion}
In this work, we investigated the nonlinear dynamics of a two-dimensional Rayleigh-B\'enard convection model as a function of the Rayleigh number and the strength of an imposed magnetic field. In the purely hydrodynamic case ($\boldsymbol{B}=0$), a bifurcation diagram was produced, showing a sequence of bifurcations previously reported by \citet{Paul2012}, where a complex route to chaos involving equilibrium points, periodic and quasiperiodic attractors, periodic traveling rolls, chaotic traveling rolls, and chaotic attractors is observed. Periodic windows, multistability, interior crises, and attractor-merging crises are observed in the absence of a magnetic field, as reported by \citet{Paul2012}. This specific route to chaos, encompassing quasiperiodic regimes, crises, and chaotic intermittent attractors, is not strictly limited to two-dimensional models; it is also observed in fully three-dimensional Rayleigh-B\'enard convection, where it strictly determines the onset threshold for intermittent dynamo action \cite{Roman2017}. In addition, we observed hysteresis in a region of coexistence of chaotic traveling rolls and fixed-point attractors.
Throughout all hydrodynamic simulations, the system exhibits only two convective vortices, which move in different ways. No vortex breaking was observed. 
%The dynamics of the system were studied using the phase space and the physical space. In the phase space, a detailed study of the system was conducted, where bifurcation diagrams were presented for different values of the control parameters used here. In the physical space, we tracked the dynamics of vortices in the magnetic and velocity fields. %Initially, an investigation of the model was carried out considering a purely hydrodynamic Rayleigh-Bénard convection (without magnetic field). Fixing the Prandtl number at $P=6.8$, we vary the reduced Rayleigh number in the interval $1 \leq r \leq 1000$. Our next step was to investigate the influence of an imposed magnetic field. To do so, we mantain the Prandtl number $P=6.8$, to the same value as in the hydrodynamic CRB case, and set the magnetic Prantl number to $P_{m}=1$. The reduced Rayleigh number was also explored for the same range ($1 \leq r \leq 1000$). The control parameter related to the magnetic field strength is the Chandrasekhar number $Q$. We explore the Chandrasekhar number in the interval $0 \leq Q \leq 50$.

The addition of a background magnetic field changes the dynamics, facilitating the appearance of chaotic traveling rolls. It also introduces the breaking of vortices in the velocity field, where the original two vortices are split into four, forming a pair of homoclinic connections that are clearly seen in the FTLE field.
The interaction between a vortex center and a saddle point causes the disappearance of one of the vortex pairs and the recovery of the two-vortex state. During this process, the magnetic field undergoes a reconnection that splits a magnetic vortex into two. Overall, the magnetic field exhibits numerous vortices that constantly interact with one another. Their interaction with the velocity field structures appears to favor transitions to traveling roll dynamics and chaos. 

\citet{laakmann2024} presented a study on 2-D magnetoconvection where a bifurcation diagram was produced as a function of the Chandrasekhar number $Q$ and compared with the non-magnetized RBC. Unlike our case, they used a square spatial domain and no-slip boundary conditions, and varied Q up to $10^3$. A sequence of pitchfork, Hopf, and saddle–node bifurcations was found. For their configuration, the use of a very high $Q$ favors the stabilization of solutions.
The intermittent transition between two-roll and three-roll structures in the velocity field was previously observed in 2-D numerical simulations of an electroconvective flow \cite{feng2026}, i.e., an RBC system subjected to a vertical electric potential difference. The use of FTLE and IVD fields, as suggested in the present work, can greatly improve the visualization of these structures and the identification of the homoclinic connections underlying these phenomena.

It is important to note that the present results are inherently restricted by the two-dimensional formulation. In fully three-dimensional environments, additional mechanisms such as vortex stretching and three-dimensional instabilities, as well as the generation of magnetic fields (prohibited in two dimensions by the Zeldovich antidynamo theorem), could significantly alter the observed transition to chaos and vortex splitting dynamics. Indeed, studies of three-dimensional magnetoconvection have already demonstrated that additional degrees of freedom, such as rotation, lead to an exceptionally rich state space with dozens of coexisting macroscopic MHD attractors and complex routes to chaos, including SNIC-like global bifurcations \cite{Roman2010}. Future three-dimensional simulations will be essential to determine exactly how these topological interactions and localized magnetic reconnections evolve within fully developed magnetohydrodynamic turbulence.

\nonumsection{Acknowledgments} \noindent This work received financial support from the Brazilian agencies CNPq (projects 405508/2021-2, 304449/2017-2 and 317522/2025-6), FAPEG (project no. 401425/2023-1), CAPES (88887.309065/2018-00) and FAPESP (2013/26258-4 and 2025/11756-6); and FCT (Portugal):  
Research Center for Systems and Technologies (UID/00147), the Associate Laboratory Advanced Production and Intelligent Systems (LA/P/0112/2020, DOI: 10.54499/LA/P/0112/2020). Part of the computations were performed on the supercomputer MareNostrum 5 (Barcelona Supercomputing Center) supported by the FCT in the framework of the computational project 2026.07585.CPCA.
 ELR would like to thank the International Space Science Institute (ISSI) in Bern, Switzerland, for the hospitality provided to the members of the team on ‘Opening new avenues in identifying coherent structures and transport barriers in the magnetised solar plasma'.

%This work had the financial support of Brazilian funding agencies CAPES (88887.309065/2018-00), CNPq (304449/2017-2) and FAPESP (2013/26258-4); Russian Scientific Foundation: project 19-11-00258 carried out in the Federal Research Center “Computer Science and Control” of the Russian Academy of Sciences, Moscow, Russian Federation; and FCT (Portugal): Base Funding - UIDB/00147/2020 and Programmatic Funding - UIDP/00147/2020 of the Systems and Technologies Center – SYSTEC, Project MAGIC – “Multi-Agent Control and Estimation for Multi-Horizon Goals Conciliation”  ref. PTDC/EEI-AUT/32485/2017, both supported by national funds through the FCT/MCTES (PIDDAC) and Project Atlantida – “Plataforma para a Monitorização do Oceano Atlântico Norte e Ferramentas para a Exploração Sustentável dos Recursos Marinhos”, Ref: NORTE-01-0145-FEDER-000040, supported by North Portugal Regional Operational Programme (NORTE 2020), under the Portugal 2020 Partnership Agreement, through the European Regional Development Fund (ERDF). A part of computations were performed on the Oblivion cluster (Evora University) in the framework of the CPCA/A0/7296/2020 project supported by FCT Advanced Computing Resources for Research and Innovation Call for Advanced Computing Projects.

%\end{multicols}

\begin{thebibliography}{38}

\bibitem[Akhtari et al.(2024)]{akhtari2024}
Akhtari, A., Zikanov, O. \& Krasnov, D. [2024] ``Magnetoconvection in a long vertical enclosure with walls of finite electrical conductivity,'' Int. J. Therm. Sci. 204, 109241, \url{https://doi.org/10.1016/j.ijthermalsci.2024.109241}.

\bibitem[Arter(1983)]{Arter1983}
Arter, W. [1983] ``Nonlinear convection in an imposed horizontal magnetic field,'' Geophys. Astrophys. Fluid Dyn. 25, 259--292, \url{https://doi.org/10.1080/03091928308221752}.

\bibitem[Bekki \& Karakisawa(1995)]{Bekki1995}
Bekki, N. \& Karakisawa, T. [1995] ``Bifurcations from periodic solution in a simplified model of two-dimensional magnetoconvection,'' Phys. Plasmas 2, 2945--2962, \url{https://doi.org/10.1063/1.871441}.

\bibitem[Bekki \& Moriguchi(2007)]{Bekki2007}
Bekki, N. \& Moriguchi, H. [2007] ``Temporal chaos in Boussinesq magnetoconvection,'' Phys. Plasmas 14, 012306, \url{https://doi.org/10.1063/1.2430517}.

\bibitem[Bekki(2025)]{yutobekki2025}
Bekki, Y. [2025] ``Impacts of small-scale dynamo on rotating columnar convection in stellar convection zones,'' Astron. Astrophys. 703, A262, \url{https://doi.org/10.1051/0004-6361/202556923}.

\bibitem[Bovino et al.(2013)]{Bovino2013}
Bovino, S., Schleicher, D. R. G. \& Schober, J. [2013] ``Turbulent magnetic field amplification from the smallest to the largest magnetic Prandtl numbers,'' New J. Phys. 15, 013055, \url{https://doi.org/10.1088/1367-2630/15/1/013055}.

\bibitem[Bushby \& Houghton(2005)]{Bushby2005}
Bushby, P. J. \& Houghton, S. M. [2005] ``Spatially intermittent fields in photospheric magnetoconvection,'' Mon. Not. R. Astron. Soc. 362, 313--320, \url{https://doi.org/10.1111/j.1365-2966.2005.09303.x}.

\bibitem[Canuto et al.(2006)]{canuto06}
Canuto, C., Hussaini, M. Y., Quarteroni, A. \& Zang, T. A. [2006] Spectral Methods: Fundamentals in Single Domains (Springer, Berlin, Heidelberg), \url{https://doi.org/10.1007/978-3-540-30726-6}.

\bibitem[Chandrasekhar(1981)]{chandra}
Chandrasekhar, S. [1981] Hydrodynamic and Hydromagnetic Stability (Dover Publications, New York).

\bibitem[Chertovskih  et al.(2010)]{Roman2010}
Chertovskih, R., Gama, S.M.A., Podvigina, O. \& Zheligovsky, V. [2010]
``Dependence of magnetic field generation by thermal convection on the rotation rate: A case study,'' Physica D: Nonlinear Phenomena,
Volume 239, Issue 13, 1188--1209,
\url{https://doi.org/10.1016/j.physd.2010.03.008}.

\bibitem[Chertovskih et al.(2015)]{Roman2015}
Chertovskih, R., Chimanski, E. V. \& Rempel, E. L. [2015] ``Route to hyperchaos in Rayleigh--B\'enard convection,'' Europhys. Lett. 112, 14001, \url{https://doi.org/10.1209/0295-5075/112/14001}.

\bibitem[Chertovskih et al.(2017)]{Roman2017} 
Chertovskih, R., Rempel, E.L. \& Chimanski, E.V. [2017] ``Magnetic field generation by intermittent convection,'' Physics Letters A,
Volume 381, Issue 38, 3300-3306, \url{https://doi.org/10.1016/j.physleta.2017.08.025}.

\bibitem[Christensen \& Aubert(2006)]{Christensen2006}
Christensen, U. R. \& Aubert, J. [2006] ``Scaling properties of convection-driven dynamos in rotating spherical shells and application to planetary magnetic fields,'' Geophys. J. Int. 166, 97--114, \url{https://doi.org/10.1111/j.1365-246X.2006.03009.x}.

\bibitem[Cioni et al.(2000)]{Cioni2000}
Cioni, S., Chaumat, S. \& Sommeria, J. [2000] ``Effect of a vertical magnetic field on turbulent Rayleigh--B\'enard convection,'' Phys. Rev. E 62, R4520--R4523, \url{https://doi.org/10.1103/PhysRevE.62.R4520}.

\bibitem[Cox \& Matthews(2002)]{cox02}
Cox, S. M. \& Matthews, P. C. [2002] ``Exponential time differencing for stiff systems,'' J. Comput. Phys. 176, 430--455, \url{https://doi.org/10.1006/jcph.2002.6995}.

\bibitem[Cukierski \& Thomas(2008)]{cukierski2008}
Cukierski, K. \& Thomas, B. [2008] ``Flow control with local electromagnetic braking in continuous casting of steel slabs,'' Metall. Mater. Trans. B 39, 94--107, \url{https://doi.org/10.1007/s11663-007-9109-3}.

\bibitem[Dawes(2007)]{Dawes2007}
Dawes, J. H. P. [2007] ``Localised convection cells in the presence of a vertical magnetic field,'' J. Fluid Mech. 570, 385--406, \url{https://doi.org/10.1017/S0022112006002795}.

\bibitem[Feng et al.(2026)]{feng2026}
Feng, Z., V\'azquez, P. A. \& Zhang, M. [2026] ``Nonlinear dynamics of electroconvective flows: From equilibria to Hopf bifurcation and chaos,'' Appl. Math. Model. 154, 116680, \url{https://doi.org/10.1016/j.apm.2025.116680}.

\bibitem[Franco et al.(2020)]{Francis2020b}
Franco, F. F., Rempel, E. L. \& Mu\~noz, P. R. [2020] ``Crisis and hyperchaos in a simplified model of magnetoconvection,'' Physica D 406, 132417, \url{https://doi.org/10.1016/j.physd.2020.132417}.

\bibitem[Goluskin et al.(2014)]{goluskin2014}
Goluskin, D., Johnston, H., Flierl, G. R. \& Spiegel, E. A. [2014] ``Convectively driven shear and decreased heat flux,'' J. Fluid Mech. 759, 360--385, \url{https://doi.org/10.1017/jfm.2014.577}.

\bibitem[Gonchenko et al.(2013)]{gonchenko13}
Gonchenko, S. V., Sim\'o, C. \& Vieiro, A. [2013] ``Richness of dynamics and global bifurcations in systems with a homoclinic figure-eight,'' Nonlinearity 26, 621--678, \url{https://doi.org/10.1088/0951-7715/26/3/621}.

\bibitem[Haller \& Yuan(2000)]{haller2000}
Haller, G. \& Yuan, G. [2000] ``Lagrangian coherent structures and mixing in two-dimensional turbulence,'' Physica D 147, 352--370, \url{https://doi.org/10.1016/S0167-2789(00)00142-1}.

\bibitem[Haller(2001)]{haller2001}
Haller, G. [2001] ``Distinguished material surfaces and coherent structures in three-dimensional fluid flows,'' Physica D 149, 248--277, \url{https://doi.org/10.1016/S0167-2789(00)00199-8}.

\bibitem[Haller(2015)]{haller2015}
Haller, G. [2015] ``Lagrangian coherent structures,'' Annu. Rev. Fluid Mech. 47, 137--162, \url{https://doi.org/10.1146/annurev-fluid-010313-141322}.

\bibitem[Haller et al.(2016)]{haller2016}
Haller, G., Hadjighasem, A., Farazmand, M. \& Huhn, F. [2016] ``Defining coherent vortices objectively from the vorticity,'' J. Fluid Mech. 795, 136--173, \url{https://doi.org/10.1017/jfm.2016.151}.

\bibitem[Hepworth(2014)]{hepworth2014}
Hepworth, B. J. [2014] ``Nonlinear two-dimensional Rayleigh--B\'enard convection,'' PhD thesis, University of Leeds, Leeds, UK.

\bibitem[Jones(2007)]{Jones2007}
Jones, C. A. [2007] ``Thermal and compositional convection in the outer core,'' Treatise on Geophysics, Vol. 8: Core Dynamics, ed. Olson, P. (Elsevier, Amsterdam), pp. 131--185, \url{https://doi.org/10.1016/B978-044452748-6/00130-9}.

\bibitem[Knobloch(1986)]{knobloch1986}
Knobloch, E. [1986] ``On convection in a horizontal magnetic field with periodic boundary conditions,'' Geophys. Astrophys. Fluid Dyn. 36, 161--177, \url{https://doi.org/10.1080/03091928608208801}.

\bibitem[Laakmann \& Boull\'e(2024)]{laakmann2024}
Laakmann, F. \& Boull\'e, N. [2024] ``Bifurcation analysis of a two-dimensional magnetic Rayleigh--B\'enard problem,'' Physica D 467, 134270, \url{https://doi.org/10.1016/j.physd.2024.134270}.

\bibitem[Lo Jacono et al.(2011)]{lojacono2011}
Lo Jacono, D., Bergeon, A. \& Knobloch, E. [2011] ``Magnetohydrodynamic convectons,'' J. Fluid Mech. 687, 595--605, \url{https://doi.org/10.1017/jfm.2011.402}.

\bibitem[Lo Jacono et al.(2012)]{lojacono2012}
Lo Jacono, D., Bergeon, A. \& Knobloch, E. [2012] ``Spatially localized magnetoconvection,'' Fluid Dyn. Res. 44, 031411, \url{https://doi.org/10.1088/0169-5983/44/3/031411}.

\bibitem[Macek \& Strumik(2010)]{macek2010}
Macek, W. M. \& Strumik, M. [2010] ``Model for hydromagnetic convection in a magnetized fluid,'' Phys. Rev. E 82, 027301, \url{https://doi.org/10.1103/PhysRevE.82.027301}.

\bibitem[McCormack et al.(2025)]{mccormack2025}
McCormack, M., Teimurazov, A., Shishkina, O. \& Linkmann, M. [2025] ``Heat transport model for the transition between scaling regimes in quasistatic and full magnetoconvection,'' Int. J. Heat Mass Transfer 241, 126641, \url{https://doi.org/10.1016/j.ijheatmasstransfer.2024.126641}.

\bibitem[Mistrangelo et al.(2021)]{mistrangelo2021}
Mistrangelo, C., B\"uhler, L., Alberghi, C., Bassini, S., Candido, L., Courtessole, C., Tassone, A., Urgorri, F. R. \& Zikanov, O. [2021] ``MHD R\&D activities for liquid metal blankets,'' Energies 14, 6640, \url{https://doi.org/10.3390/en14206640}.

\bibitem[Mondal et al.(2018)]{Mondal2018}
Mondal, H., Das, A. \& Kumar, K. [2018] ``Onset of oscillatory Rayleigh--B\'enard magnetoconvection with rigid horizontal boundaries,'' Phys. Plasmas 25, 012119, \url{https://doi.org/10.1063/1.5009540}.

\bibitem[M{\"u}ller(1988)]{muller1988}
M{\"u}ller, G. [1988] ``Convection and inhomogeneities in crystal growth from the melt,'' Crystal Growth from the Melt, Crystals, Vol. 12 (Springer, Berlin, Heidelberg), pp. 1--136, \url{https://doi.org/10.1007/978-3-642-73208-9_1}.

\bibitem[Paul et al.(2012)]{Paul2012}
Paul, S., Verma, M. K., Wahi, P., Reddy, S. K. \& Kumar, K. [2012] ``Bifurcation analysis of the flow patterns in two-dimensional Rayleigh--B\'enard convection,'' Int. J. Bifurcation Chaos 22, 1230018, \url{https://doi.org/10.1142/S0218127412300182}.

\bibitem[Podvigina (2008)]{Podv2008}
Podvigina, O. M. [2008] ``Magnetic field generation by convective flows in a plane layer: the dependence on the Prandtl numbers,” Geophysical \& Astrophysical Fluid Dynamics 102 (4): 409--33, \url{https://doi.org/10.1080/03091920701841945}.

\bibitem[Podvigina et al.(2015)]{Olga2015}
Podvigina, O., Zheligovsky, V., Rempel, E. L., Chian, A. C.-L., Chertovskih, R. \& Mu\~noz, P. N. [2015] ``Two-parameter bifurcation study of the regularized long-wave equation,'' Phys. Rev. E 92, 032906, \url{https://doi.org/10.1103/PhysRevE.92.032906}.

\bibitem[Proctor \& Weiss(1982)]{Proctor1982}
Proctor, M. R. E. \& Weiss, N. O. [1982] ``Magnetoconvection,'' Rep. Prog. Phys. 45, 1317--1379, \url{https://doi.org/10.1088/0034-4885/45/11/003}.

\bibitem[Rayleigh(1916)]{rayleigh1916}
Rayleigh, Lord [1916] ``On convection currents in a horizontal layer of fluid, when the higher temperature is on the under side,'' Philos. Mag. J. Sci. 32, 529--546, \url{https://doi.org/10.1080/14786441608635602}.

\bibitem[Reddy(2020)]{Reddy2020}
Reddy, M. K. [2020] ``Numerical simulation of Rayleigh--B\'enard convection in an inclined enclosure under the influence of magnetic field,'' J. King Saud Univ. Sci. 32, 486--495, \url{https://doi.org/10.1016/j.jksus.2018.07.010}.

\bibitem[Rempel et al.(2013)]{rempel2013}
Rempel, E. L., Chian, A. C.-L., Brandenburg, A., Mu\~noz, P. R. \& Shadden, S. C. [2013] ``Coherent structures and the saturation of a nonlinear dynamo,'' J. Fluid Mech. 729, 309--329, \url{https://doi.org/10.1017/jfm.2013.290}.

\bibitem[Rempel et al.(2016)]{rempel2016}
Rempel, E. L., Chian, A. C.-L., Beron-Vera, F. J., Szanyi, S. \& Haller, G. [2016] ``Objective vortex detection in an astrophysical dynamo,'' Mon. Not. R. Astron. Soc. Lett. 466, L108--L112, \url{https://doi.org/10.1093/mnrasl/slw248}.

\bibitem[Rempel et al.(2019)]{rempel2019}
Rempel, E. L., Gomes, T. F. P., Silva, S. S. A. \& Chian, A. C.-L. [2019] ``Objective magnetic vortex detection,'' Phys. Rev. E 99, 043206, \url{https://doi.org/10.1103/PhysRevE.99.043206}.

\bibitem[Rucklidge \& Matthews(1996)]{rucklidge1996}
Rucklidge, A. M. \& Matthews, P. C. [1996] ``Analysis of the shearing instability in nonlinear convection and magnetoconvection,'' Nonlinearity 9, 311--351, \url{https://doi.org/10.1088/0951-7715/9/2/003}.

\bibitem[Schaeffer et al.(2017)]{Schaeffer2017}
Schaeffer, N., Jault, D., Nataf, H.-C. \& Fournier, A. [2017] ``Turbulent geodynamo simulations: A leap towards Earth's core,'' Geophys. J. Int. 211, 1--29, \url{https://doi.org/10.1093/gji/ggx265}.

\bibitem[Stoica \& Moses(2005)]{Stoica2005}
Stoica, P. \& Moses, R. [2005] Spectral Analysis of Signals (Prentice Hall, Upper Saddle River, New Jersey).

\bibitem[Thual(1992)]{Thual1992}
Thual, O. [1992] ``Zero-Prandtl-number convection,'' J. Fluid Mech. 240, 229--258, \url{https://doi.org/10.1017/S0022112092000089}.

\bibitem[Wang et al.(2020)]{wang2020}
Wang, Q., Chong, K. L., Stevens, R. J. A. M., Verzicco, R. \& Lohse, D. [2020] ``From zonal flow to convection rolls in Rayleigh--B\'enard convection with free-slip plates,'' J. Fluid Mech. 905, A21, \url{https://doi.org/10.1017/jfm.2020.793}.

\bibitem[Weiss \& Proctor(2014)]{Weiss2014}
Weiss, N. O. \& Proctor, M. R. E. [2014] Magnetoconvection (Cambridge University Press, Cambridge), \url{https://doi.org/10.1017/CBO9780511667459}.

\bibitem[Wilczynski et al.(2019)]{wilczynski2019}
Wilczynski, F., Hughes, D. W., Van Loo, S., Arter, W. \& Militello, F. [2019] ``Stability of scrape-off layer plasma: A modified Rayleigh--B\'enard problem,'' Phys. Plasmas 26, 022510, \url{https://doi.org/10.1063/1.5064765}.

\bibitem[Winchester et al.(2022)]{winchester2022}
Winchester, P., Howell, P. D. \& Dallas, V. [2022] ``The onset of zonal modes in two-dimensional Rayleigh--B\'enard convection,'' J. Fluid Mech. 939, A8, \url{https://doi.org/10.1017/jfm.2022.185}.

\bibitem[Yan et al.(2019)]{Yan2019}
Yan, M., Calkins, M. A., Maffei, S., Julien, K., Tobias, S. M. \& Marti, P. [2019] ``Heat transfer and flow regimes in quasi-static magnetoconvection with a vertical magnetic field,'' J. Fluid Mech. 877, 1186--1206, \url{https://doi.org/10.1017/jfm.2019.615}.


\end{thebibliography}
\end{document}